\documentclass[prd,aps,superscriptaddress,amssymb,amsmath,amsfonts]{revtex4}
\usepackage{latexsym}
\usepackage[latin1]{inputenc}
\usepackage[dvips]{graphicx}
\usepackage{color}

\newcommand{\da}{\dagger}

\newcommand{\be}{\begin{equation}} 
\newcommand{\ee}{\end{equation}} 
\newcommand{\ba}{\begin{eqnarray}} 
\newcommand{\ea}{\end{eqnarray}}

\newcommand{\pa}{\partial}

\usepackage{dcolumn}

\begin{document}
                                                                                
\title{$D$-meson propagation in hot dense matter}
\author{Laura Tolos}
\affiliation{Institut de Ci\`encies de l'Espai (IEEC/CSIC), Campus Universitat Aut\`onoma
de Barcelona, Facultat de Ci\`encies, Torre C5, E-08193 Bellaterra, Spain}
\affiliation{Frankfurt Institute for Advances Studies. Johann Wolfgang Goethe University, Ruth-Moufang-Str. 1,
60438 Frankfurt am Main, Germany }
\author{Juan M. Torres-Rincon}
\affiliation{Institut de Ci\`encies de l'Espai (IEEC/CSIC), Campus Universitat Aut\`onoma
de Barcelona, Facultat de Ci\`encies, Torre C5, E-08193 Bellaterra, Spain}
\date{\today}

\begin{abstract}
The drag force and diffusion coefficients for a $D$ meson are calculated in hot dense matter composed of light mesons and baryons, such as formed in heavy-ion collisions.  We use a unitarized approach based on effective models for the interaction of a $D$ meson with hadrons, which are
compatible with chiral and heavy quark symmetries.  We study the propagation of the $D$ meson in the hadron matter in two distinct cases. On the one
hand, we analyze the propagation of $D$ mesons in matter at vanishing baryochemical potential $\mu_B$, which is relevant for high-energetic
collisions at LHC or RHIC. On the other hand, we show the propagation of $D$ mesons in the hadronic medium following isentropic trajectories, appropriate at
FAIR and NICA heavy-ion experiments. We find a negligible baryon contribution to the transport coefficients at $\mu_B=0$. However at $\mu_B >0$ we obtain a
large correction to the transport coefficients with the inclusion of nucleons and $\Delta$ baryons. 
The relaxation time for $D$ mesons is reduced by a factor 2--3 in the latter case, producing a more thermalized $D$-meson spectrum for FAIR physics than for the
typical LHC energies.  We finally present results for the spatial diffusion coefficient of a $D$ meson in hadronic matter and the possible existence of a
minimum near the phase transition to the quark-gluon plasma at zero and finite baryochemical potential.
\end{abstract}

\maketitle

\section{Introduction}

Many of the hadronic studies that have been developed within the context of heavy-ion collisions are devoted to the extraction
of properties of the deconfined phase, the quark-gluon plasma (QGP). This phase of QCD is produced after the rapid thermalization process which follows the nucleus-nucleus collision.
The search of some QGP signature imprinted in the hadronic phase becomes of vital importance if one wants to study the primordial phase from what is actually tracked in the detector.

   Some of the probes that carry information from the initial stages of the expansion are, for example, direct photons~\cite{Adare:2008ab,Wilde:2012wc} and the jet quenching of high-$p_T$
hadrons~\cite{Adcox:2001jp}. One of the cleanest hadronic probes of the post-thermalization stage are heavy quarks. Due to their large mass ($m_c \simeq 1.3$ GeV, $m_b \simeq 4.3$ GeV)
in comparison to the mass of the light-flavor quarks, they have large relaxation times and, therefore, they cannot totally relax during the fireball's expansion
(around ten Fermi for Pb-Pb collisions at the LHC~\cite{Aamodt:2011mr}). For these reasons, the heavy mesons (once the $c$ and $b$ quarks have hadronized) are
nowadays attracting much interest in the hadronic physics community.

The electrons coming from the heavy-meson semileptonic decays are taken as probes of the heavy-meson dynamics, e.g. the heavy-meson drag and diffusion are usually
analyzed through the nuclear suppression factor and elliptic flow of these electrons. Examples of these observables at nearly vanishing baryochemical potential can
be found at both the RHIC~\cite{Adare:2006nq} and the LHC~\cite{Masciocchi:2011fu}. However, it is also possible to reconstruct heavy mesons from the decay products that
are seen in the detector. Focusing on the charm sector, to which the present article is devoted, the nuclear suppression factor has already been extracted for $D$ mesons
 by the ALICE Collaboration in Ref.~\cite{ALICE:2012ab}. Also the elliptic flow $v_2$ has recently been obtained by the same collaboration~\cite{delValle:2012qw}.

Analogously to the case of the viscosities for the light particles, these observables are strongly influenced by
the transport coefficients of the hadronic medium. The present paper focuses on the theoretical calculation
of the $D$-meson transport coefficients at finite temperature and baryonic chemical potential. In particular, we provide a consistent method
to calculate the drag force and diffusion coefficients of $D$ mesons from effective field theories.

The use of an effective theory incorporating both chiral and heavy-quark symmetries to calculate
the diffusion coefficients was first implemented in Ref.~\cite{Laine:2011is}. However, the validity of the results in that paper
is quite limited in temperature, roughly between $30$ MeV $<T< 80$ MeV.
In Ref.~\cite{He:2011yi} the drag and diffusion coefficients of $D$ mesons were obtained using parametrized interactions with
light mesons and baryons. Although their scattering amplitudes are only valid close to the resonant states, the approach captures
the correct behavior of cross sections and provides a fair estimation of the transport coefficients. In Ref.~\cite{Ghosh:2011bw} effective Lagrangians were used
to obtain the scattering amplitudes of $D$ mesons with light mesons and baryons at leading order. In this case, the scattering amplitudes rapidly grow with energy and break
the unitarity condition for the scattering $S$ matrix. Therefore, the transport coefficients turn out to be unphysically large. Finally, in Ref.~\cite{Abreu:2011ic} the Lagrangian for the interaction
 of $D$ mesons with pions at next-to-leading order was developed together with the implementation of a unitarization method. The unitarization procedure provides a realistic interaction
 between $D$ mesons and pions up to temperatures of the order of the pion mass. A similar effective Lagrangian with an extension to kaons and $\eta$ mesons was used
for the bottom sector in Ref.~\cite{Abreu:2012et}.

In this paper, we obtain the drag and diffusion coefficients for a $D$ meson propagating in a meson gas which contains the complete set of pseudo-Goldstone bosons (pions, kaons
and $\eta$ mesons) as well as baryonic degrees of freedom (nucleons and $\Delta$ baryons). Thus, we 
extend the results of Ref.~\cite{Abreu:2011ic} to include the next relevant degrees of freedom.
All the scattering amplitudes are obtained from effective Lagrangians consistent with chiral and heavy-quark spin symmetries. 
Additionally, the amplitudes are unitarized to restore the $S$-matrix unitarity, lost due to the truncation of the perturbative expansion.
This leads to the appearance of resonant states which are implicitly taken into account in the interaction's description.
This fact allows to avoid unphysical amplitudes and have a well-controlled energy dependence, essential for the description of the transport coefficients.

We first present results for the transport coefficients at zero baryochemical potential, i.e  $\mu_B=0$. The calculation of the transport coefficients for $\mu_B=0$ is
relevant for high-energetic collisions, such as those at the LHC or RHIC (at its top colliding energies). Next, we show our results for $\mu_B>0$ trajectories within the
QCD phase diagram that are obtained by fixing the entropy per baryon, i.e. fixed $s/n_B$. The future heavy-ion collision
experiments at FAIR or NICA facilities will explore the phase diagram at finite temperature and baryochemical potential, thus allowing
to analyze the propagation of $D$ mesons in a hot dense baryonic environment. As we shall see in the following, in contrast to the $\mu_B=0$ case, the contribution
of the $D$-meson scattering with nucleons to the transport coefficients is significant.  Finally, we study the behavior of the spatial diffusion coefficient in the
hadronic phase for $\mu_B=0$, as well as for isentropic trajectories, and present a few calculations in the quark-gluon plasma domain in both cases. We analyze
the possibility of a minimum of the spatial diffusion coefficient at the deconfinement phase transition, in particular at finite baryochemical potential.

The structure of the paper is as follows. In Sec.~\ref{sec:transport} we  introduce the transport coefficients
relevant for heavy meson dynamics. The details on the $D$-meson interactions with light mesons and baryons are given in Sec.~\ref{sec:interaction},
where the effective Lagrangian and the unitarization method are described for both meson and baryon sectors.
We present our results in Sec.~\ref{sec:results}, where the drag force and the diffusion coefficients are displayed for zero baryochemical potential as well as for
isentropic trajectories. In Sec.~\ref{sec:dif} we show results for the spatial diffusion coefficient in the baryonic and quark-gluon plasma phases at zero
and finite baryochemical potential, analyzing the possible existence of a minimum in the spatial diffusion coefficient in the phase transition. Our conclusions and outlook are given in Sec.~\ref{sec:conclusions}.

\section{Drag and diffusion coefficients for $D$ mesons \label{sec:transport}}

The momentum-space distribution of $D$ mesons out of equilibrium obeys the Boltzmann equation \cite{landau1981course}. The Boltzmann equation reduces, however, to a
 much simpler expression, the Fokker-Planck equation, in the limit where the mass of the particle propagating in the thermal bath is much greater than the mass 
of the surrounding particles as well as the temperature of the heat bath. Thus, the momentum-space distribution function of $D$ mesons with momentum $\bf{p}$, $f(t,{\bf p})$, in a gas of light particles satisfies the Fokker-Planck equation
\be \frac{\pa f(t,{\bf p})}{\pa t} = \frac{\pa}{\pa p_i} \left\{ F_i (\mathbf{p}) f(t,{\bf p}) + \frac{\pa}{\pa p_j} \left[ \Gamma_{ij} (\mathbf{p}) f(t,\bf{p}) \right] \right\} \ ,\ee
with $i,j=1,2,3$ the spatial indices. The microscopical expressions for the quantities $F_i$ and $\Gamma_{ij}$ result from the matching with the Boltzmann equation~\cite{landau1981course,Abreu:2011ic} and are given by
\be F_i (\mathbf{p})= \int d\mathbf{k} \ w(\mathbf{p},\mathbf{k}) \ k_i \ , \ee
\be \Gamma_{ij} (\mathbf{p})= \frac{1}{2} \int d\mathbf{k} \ w(\mathbf{p},\mathbf{k}) \ k_i k_j \ , \ee
where $w(\mathbf{p},\mathbf{k})$ is the collision rate for a $D$ meson with initial and final momenta, $\bf{p}$  and $\bf{p-k}$, respectively, being $\mathbf{k}$ the transferred momentum. We observe that $F_i$ behaves as a friction term or drag force representing the average momentum change of the $D$ meson whereas $\Gamma_{ij}$ acts as a diffusion coefficient in momentum space, forcing a broadening of the average momentum distribution of the $D$ meson. The collision rate $w({\bf p},{\bf k})$ reads

\ba w(\mathbf{p},\mathbf{k})  &=&  g_l \int \frac{d^3q}{(2\pi)^9} \ n_{F,B} (E_l(q),T) \ \left[ 1\pm n_{F,B} (E_l(q+k),T) \right]  \frac{1}{2E_D(p)} \frac{1}{2E_l(q)} 
\frac{1}{2E_D(p-k)} \frac{1}{2E_l(q+k)} \nonumber \\ 
& \times & (2\pi)^4 \delta (E_D(p)+E_l(q)-E_D(p-k)-E_l(q+k)) \ \overline{|\mathcal{M}^2|} \ , 
\label{omega}
\ea
where $D$ labels the charmed meson and $l$ represents the light hadron belonging to the thermal bath. The quantity $g_l$ stands for the spin-isospin
degeneracy factor of the light hadron while the light hadron bath is taken to be at equilibrium with distribution function $n_{F,B} (E_l,T)$, for
 Fermi-Dirac or Bose-Einstein particles, respectively. Indeed, for mesons we include the Bose enhancement factor $\left[1+n_B(E_l,T)\right]$ whereas for baryons
we consider the Pauli blocking term $\left[ 1-n_F(E_l,T) \right]$.  The invariant scattering matrix element is given by $\mathcal{M}$, which
contains the microscopical details of the collision.

Assuming an isotropic bath, the transport coefficients $F_i (\mathbf{p})$ and $ \Gamma_{ij} (\mathbf{p})$ can be written in terms of three scalar functions as
\ba F_i (\mathbf{p}) & = & F(p) \  p_i \ , \\
\Gamma_{ij} (\mathbf{p}) &= & \Gamma_0 (p) \left( \delta_{ij} - \frac{p_i p_j}{p^2}\right) + \Gamma_1 (p) \ \frac{p_i p_j}{p^2} \ . 
\ea
The explicit expressions for $F(p)$, $\Gamma_0(p)$ and $\Gamma_1(p)$  are then obtained as a function of $w(\mathbf{p},\mathbf{k})$:
\ba \label{eq:Fcoeff}
F(p) & =& \int d\mathbf{k} \ w(\mathbf{p},\mathbf{k}) \frac{k_ip^i}{p^2} \ , \\
\label{eq:G0coeff} \Gamma_0(p) & =& \frac{1}{4} \int d\mathbf{k} \ w(\mathbf{p},\mathbf{k}) \left[ \mathbf{k}^2
- \frac{(k_ip^i)^2}{p^2} \right] \ ,  \\
\label{eq:G1coeff} \Gamma_1(p) & =& \frac{1}{2} \int d\mathbf{k} \ w(\mathbf{p},\mathbf{k}) \frac{(k_ip^i)^2}{p^2} \ . 
\ea
where the dynamics comes through the collision rates and, hence, via the invariant scattering matrix elements $\mathcal{M}$. In the following section we present the details of the interaction of a $D$ meson with hadrons in a thermal bath.

\section{$D$ mesons in a hadronic thermal bath\label{sec:interaction}}

In order to obtain the invariant matrix elements $\mathcal{M}$ of Eq.~(\ref{omega}), we need to evaluate the scattering amplitudes $T$ for the interaction of $D$ mesons with light mesons and baryons. In both meson and baryon sectors, this amplitude follows the standard multichannel scattering (integral) Bethe-Salpeter equation,
\be
T=\ V \ +\ V \ G\ T \ ,
\label{bs}
\ee
where $V$ is the potential resulting from the meson-meson (meson-baryon) effective Lagrangian and $G$ is the two-particle meson-meson (meson-baryon) propagator. 

The kernel $V$ is a matrix that consists of all possible meson-meson (meson-baryon) transitions. We focus on the interaction of $D$ mesons with the
pseudo-Goldstone bosons ($\pi$, $K$, $\bar K$ and $\eta$) as well as with the lightest baryons ($N$ and $\Delta$). We make use of the effective
 model of Ref.~\cite{Abreu:2011ic} for the interaction of $D$ mesons with light mesons, which is consistent with chiral symmetry and heavy-quark spin symmetry (HQSS). For
 the scattering of $D$ mesons with  baryons, we take into account two different schemes: the SU(4) Weinberg-Tomozawa (WT) interaction model of
 Ref.~\cite{Mizutani:2006vq} and the SU(6)$\times$HQSS WT scheme of Refs.~\cite{GarciaRecio:2008dp,Romanets:2012hm}. Similarly to the meson-meson sector, both meson-baryon models fulfill chiral symmetry in the light-quark sector while heavy-quark spin symmetry constraints are respected in the heavy-quark sector. The main features of all these models will be specified in the next subsections.

The $V$ kernel can be factorized in the on-mass shell~\cite{Oller:1997ti}, so the scattering amplitudes $T$ of Eq.~(\ref{bs}) are the solutions of a set of linear algebraic coupled equations 
\be
T_{ij}=[1 -  V G ]_{ik}^{-1} \ V_{kj} \ ,
\ee
where $i$ and $j$ indicate the initial meson-meson (meson-baryon) and final meson-meson (meson-baryon) systems, respectively. This approach is practically equivalent to
the so-called $N/D$ method \cite{Oller:2000fj}. In the on-shell ansatz, the two-particle propagators ---often called loop functions--- form a diagonal matrix $G$. The loop
functions read
\be \label{eq:loop} G_r (\sqrt{s})= i \gamma_r \int \frac{d^4q}{(2\pi)^4} \frac{1}{(P-q)^2-M_r^2+i\epsilon} \frac{1}{q^2-m_r^2+i\epsilon} \ , \ee
with the total four-momentum $P$ related to the center-of-mass squared energy $s$ by $s=P^2$,  and $q$ being the relative four-momentum in the center-of-mass frame. 
The quantities $m_r$ and $M_r$ stand for the masses of the two particles propagating in the intermediate channel $r$, i.e,  two mesons or a meson and a baryon. The
factor $\gamma_r$ has been introduced to account for possible different normalization of the meson-meson and meson-baryon interactions. In fact, as we will see
in the following subsections, $\gamma_r=1$ for the adimensional meson-meson $V$ kernel while for the meson-baryon sector $\gamma_r = 2 M_r$, with $M_r$ being the mass
of the baryon. The meson-meson (meson-baryon) loop functions are divergent and are regularized by means of dimensional regularization. 
 
Once the scattering amplitudes $T_{ij}$ are computed, the invariant matrix elements $\mathcal{M}_{ij}$ are given by
\ba
\mathcal{M}_{ij} (\sqrt{s})& = & \gamma_i^{1/2} \gamma_j^{1/2} \ T_{ij} (\sqrt{s}) \ . 
\ea

\subsection{$D$-meson interaction with light mesons}

The charm degree of freedom has recently been incorporated in meson-meson models~\cite{Kolomeitsev:2003ac,Hofmann:2003je,Guo:2006fu,dany,Molina:2008nh,danyax,Branz:2009yt,Faessler:2007gv,
Segovia:2008zz,FernandezCarames:2009zz,Gutsche:2010zza,HidalgoDuque:2012pq,Guo:2008zg} in order to study the nature of many of the observed states with hidden charm and
open charm. In particular, the chiral Lagrangian density that we will use to describe the interaction between the spin-zero and spin-one $D$ mesons and pseudoscalar Goldstone bosons reads
\be \mathcal{L} = \mathcal{L}_{LO} + \mathcal{L}_{NLO} \ , \ee
where LO and NLO refer to the leading order and next-to-leading order in the chiral expansion, always keeping leading order in the heavy-quark expansion. The LO contribution is  given by~\cite{Lutz:2007sk,Guo:2008gp,Guo:2009ct,
Geng:2010vw,Abreu:2011ic}
\ba \mathcal{L}_{LO} & = & \langle \nabla^\mu D \nabla_\mu D^\dag \rangle - m_D^2 \langle DD^{\dag} \rangle 
- \langle \nabla^\mu D^{*\nu} \nabla_\mu D^{*\dag}_{\nu} \rangle + m_D^2 \langle D^{*\mu} D_\mu^{* \dag} \rangle \nonumber \\
 &+& ig \langle D^{* \mu} u_\mu D^\dag - D u^\mu D_\mu^{*\dag} \rangle + \frac{g}{2m_D} \langle D^*_\mu u_\alpha \nabla_\beta D_\nu^{*\dag} -
\nabla_\beta D^*_\mu u_\alpha D_\nu^{*\dag}\rangle  \epsilon^{\mu \nu \alpha\beta}\ , \ea
where $D=(D^0,D^+,D_s^+)$  and $D_{\mu}^*=(D^{*0},D^{*+},D_s^{*+})_{\mu}$ are the SU(3) antitriplets of spin-zero and spin-one $D$ mesons with the chiral limit
mass $m_D$, respectively, while the brackets denote the trace in flavor space. The LO Lagrangian contains the kinetic and mass terms of the
$D$ and $D^*$ mesons as well as two interaction terms. As done in Ref.~\cite{Abreu:2011ic}, we have used HQSS to relate the two interaction terms using the same coupling constant $g$. The axial vector field is 
\be u_\mu = i \left( u^\dag \pa_\mu u - u \pa_\mu u^\dag \right) \ , \ee
whereas the covariant derivative is defined as
\be \nabla_\mu = \pa_\mu - \frac{1}{2} \left( u^\dag \pa_\mu u + u \pa_\mu u^\dag \right) \ , \ee
with $u=\sqrt{U}$ being the exponential matrix including all Goldstone bosons 
\be U=\exp \left(\frac{\sqrt{2} i\Phi}{f_\pi}  \right) \ , \ee
with
\be \Phi= \left( 
\begin{array}{ccc}
\frac{1}{\sqrt{2}} \pi^0 + \frac{1}{\sqrt{6}} \eta  & \pi^+ & K^+ \\
\pi^- & - \frac{1}{\sqrt{2}} \pi^0 + \frac{1}{\sqrt{6}} \eta & K^0 \\
K^- & {\bar K}^0 & - \frac{2}{\sqrt{6}} \eta 
\end{array}
\right) \ , \ee
and $f_\pi$ being the Goldstone boson decay constant in the chiral limit, which we take to be $f_\pi=93$ MeV.

The NLO chiral Lagrangian reads \cite{Lutz:2007sk,Guo:2008gp,Guo:2009ct,
Geng:2010vw,Abreu:2011ic}
 \ba
\mathcal{L}^{NLO} & = & - h_0  \langle D D^{\da} \rangle \langle \chi_+ \rangle  + h_1  \langle D \chi_+ D^{\da} \rangle + h_2  \langle D D^{\da} \rangle
\langle u^{\mu}u_{\mu} \rangle  \nonumber  + h_3  \langle D u^{\mu}u_{\mu} D^{\da} \rangle + h_4  \langle \nabla _{\mu} D  \,  \nabla _{\nu} D^{\da} \rangle \langle u^{\mu}u^{\nu} \rangle \nonumber \\
& & + h_5  \langle \nabla _{\mu} D   \{ u^{\mu}, u^{\nu} \} \nabla _{\nu}D^{\da} \rangle + \tilde{h}_0  \langle D^{\ast \mu} D^{\ast \da}_{\mu} \rangle \langle \chi_+ \rangle 
 - \tilde{h}_1  \langle D^{\ast \mu} \chi_+ D^{\ast \da}_{\mu} \rangle   - \tilde{h}_2 \langle  D^{\ast \mu}  D^{\ast \da}_{\mu} \rangle \langle u^{\nu} u_{\nu} \rangle \nonumber \\
& &  - \tilde{h}_3 \langle  D^{\ast \mu}  u^{\nu}u_{\nu} D^{\ast \da}_{\mu} \rangle - \tilde{h}_4  \langle \nabla _{\mu}D ^{\ast \alpha}  \,  \nabla _{\nu}D^{\ast \da}_{\alpha} \rangle \langle u^{\mu}u^{\nu} \rangle  
 - \tilde{h}_5 \langle \nabla _{\mu}D ^{\ast \alpha}  \{ u^{\mu}, u^{\nu} \}\nabla _{\nu}D^{\ast \da} _{\alpha} \rangle \ ,
\label{lag2}
\ea
where
\be
\chi _{+} =  u^{\da} \chi u^{\da} +u \chi u \ ,
\label{u_chi}
\ee
with $\chi = \mathrm{diag}(m^2_{\pi}, m^2_{\pi}, 2 m^2_{K} -m^2_{\pi})$ being the mass matrix. The NLO contribution contains twelve
low-energy constants (LECs), $h_i$ and $\tilde{h}_i (i=0,...,5)$, which need to be fixed. The number of free LECs can be reduced, though, working at LO in HQSS,
where $\tilde{h}_i=h_i$, and keeping lowest order in $N_c$ counting so that one only needs to consider odd LECs.

If we keep LO  in the heavy-quark mass expansion, the final expression for the tree-level scattering amplitude of a $D$ meson scattered with a light meson  reads  \cite{Abreu:2011ic} 
\be V^{IJSC}= \frac{C_0}{4 f_\pi^2} (s-u) + \frac{2C_1 h_1}{f_\pi^2} + \frac{2C_2}{f_\pi^2} h_3 (p_2 \cdot p_4) + \frac{2C_3}{f_\pi^2} h_5 [(p_1 \cdot p_2) (p_3 \cdot p_4)
+ (p_1 \cdot p_4) (p_2 \cdot p_3)] \ , \ee
where $p_1$ and $p_2$ are the momenta of the incoming hadrons, $p_3$ and $p_4$ are the outgoing momenta, $s=(p_1 + p_2)^2$ and $u=(p_1 - p_4)^2$. The labels $I,J,S$ and $C$ denote the channel's isospin, spin, 
strangeness and charm quantum number, respectively (we restrict ourselves to $C=1$). The first LEC is fixed to $h_1=-0.41$ using the mass difference
between the $D$ and $D_s$ mesons~\cite{Geng:2010vw}, whereas $h_3$ and $h_5$ are the two free LECs. The quantities $C_i$ are the isospin coefficients of the different
scattering amplitudes of $D$ mesons with $\pi$, $K$, $\bar K$ and $\eta$ mesons, as shown in Table~\ref{tab:isoscoeff}.  

\begin{table}[h]
\begin{ruledtabular}
\begin{tabular}{ccccccccc}
$C_{i}$ &  $D \pi(\frac{1}{2})$ & $D\pi(\frac{3}{2})$ & $D \bar{K} (0)$ & $D \bar{K}(1)$
 & $D K(0)$ & $D K(1)$ & $D\eta (\frac{1}{2})$ & $D\pi \leftrightarrow D\eta (\frac{1}{2})$ \\
\hline
$C_0 $  & -2 & 1  & -1  & 1  &  -2  & 0 & 0 &0     \\
$C_1 $  & $ - m_{\pi} ^2$ & $ - m_{\pi} ^2$ & $  m_{K} ^2$ & $ - m_{K} ^2$ 
& $ - 2 m_{K} ^2$ & 0 & $ - m^2_\pi/3$ & $-m^2_\pi$ \\
$C_2 $  & 1 & 1 & -1 & 1 & 2 & 0 & 1/3  &  1    \\
$C_3 $  & 1 & 1 & -1 & 1 & 2 & 0 & 1/3  &  1   \\
\end{tabular}
\end{ruledtabular}
\caption{ \label{tab:isoscoeff} Isospin coefficients of the scattering amplitudes for the $D$ meson--light meson channels with total isospin $I$. }
\end{table}

Compared to Ref.~\cite{Abreu:2011ic}, we consider not only the scattering of $D$ mesons with $\pi$ but also with $K$, $\bar K$ and $\eta$ mesons, in an analogous way to what was
done in the bottom sector in Ref.~\cite{Abreu:2012et}. Moreover, we take into account the nonvanishing inelastic amplitude $D\pi \leftrightarrow D\eta$. Note that this
mixing only occurs at NLO because its $C_0=0$, and, therefore, it was neglected in Ref.~\cite{Abreu:2012et} due to its small effect. However, in this work we also
incorporate the effect of the ($D \pi$, $D \eta$) coupled-channel structure when unitarizing the amplitude. 
 We keep the same  subtraction point for the three channels but we slightly change the values of the LECs with respect to those
in Ref.~\cite{Abreu:2011ic} in order to fix the pole position and width of the $D_0 (2400)$ resonance. The procedure of fixing the LECs is extensively
detailed in Ref.~\cite{Abreu:2011ic}. The values of the LECs used in this work are $h_3=5.5$ and $h_5=-0.45$ GeV$^{-2}$, which are in agreement with the estimate from
lattice-QCD data in Ref.~\cite{Liu:2012zya}.

\subsection{$D$-meson interaction with $N$ and $\Delta$}
\label{dn}

Approaches based on coupled-channel dynamics have recently been constructed in the meson-baryon sector with charm  
\cite{Tolos:2004yg, Tolos:2005ft, Lutz:2003jw, Lutz:2005ip, Hofmann:2005sw,Hofmann:2006qx, Lutz:2005vx, Mizutani:2006vq, Tolos:2007vh,
JimenezTejero:2009vq, Haidenbauer:2007jq, Haidenbauer:2008ff,Haidenbauer:2010ch, Wu:2010jy, Wu:2010vk, Wu:2012md, Oset:2012ap}, partially
motivated by the parallelism between the $\Lambda(1405)$ and the $\Lambda_c(2595)$. In this work we consider two coupled-channel schemes that have proven to be
successful in describing existing experimental charmed states. On one hand, we make use of a model based on the dominance of $t$-channel vector meson exchange for the
$s$-wave interaction between $D$ mesons and baryons of Ref.~\cite{Mizutani:2006vq}. On the other hand, we take into account a recent unitarized coupled-channel scheme
that explicitly implements HQSS in the charm sector~\cite{GarciaRecio:2008dp,Romanets:2012hm,Garcia-Recio:2013gaa}.

\subsubsection{SU(4) Weinberg-Tomozawa model}
\label{su4}

The first transition potential for $D$ mesons with baryons that we use is based on a type of broken SU(4) $s$-wave WT interaction, which results from
the $t$-channel vector meson exchange interaction in the $t \rightarrow 0$ limit \cite{Mizutani:2006vq}. The interaction kernel is given by 
\be 
\label{eq:pert_amp}
 V_{ij}^{IJSC}= \frac{\chi_c \ {D}^{IJSC}_{ij}}{4f^2} \ (2 \sqrt{s} -M_i -M_j ) \sqrt{\frac{M_i+E_i}{2M_i} }  \sqrt{ \frac{M_j+E_j}{2M_j} }  \ ,
 \ee
where $i,j$ denote the initial and final meson-baryon states formed by a pseudoscalar meson and a $1/2^+$ baryon. The quantities $M_i$ and $E_i$ are the mass and energy of
the baryon in a certain channel $i$, respectively, and $\sqrt{s}$ is the center-of-mass energy,  while the weak decay constant is fixed to $f=1.15 f_{\pi}$. The
 structure coefficients for SU(4) symmetry are given by ${D}^{IJSC}_{ij}$ (see \cite{Mizutani:2006vq} for the exact values). The SU(4) symmetry
 is severely broken in nature, so we implement a symmetry-breaking mechanism. The breaking comes through the physical hadron masses as well as by means of the factor $\chi_c$ which indicates the suppression of the charm-exchange transitions. 

We are interested in the transitions involving $D$ mesons with nucleons. Thus, we focus on the nonstrange ($S = 0$) and singly charmed ($C = 1$) sector which includes
 $\pi \Sigma_c$, $DN$,$\eta \Lambda_c$, $K \Xi_c$, $K \Xi_c^\prime$, $D_s \Lambda$ and $\eta^\prime \Lambda_c$ for $I=0,J=1/2$; and $\pi \Lambda_c$, $\pi \Sigma_c$, $DN$,
$K \Xi_c$, $\eta \Sigma_c$, $K \Xi^\prime_c$, $D_s \Sigma$ and $\eta^\prime \Sigma_c$ for $I=0,J=3/2$.

Given the transition potentials within the SU(4) WT scheme, we can now solve the on-shell Bethe-Salpeter equation in coupled channels so as to calculate the scattering amplitudes. The loop function
is regularized using dimensional regularization so as to generate dynamically the $I=0$ $\Lambda_c(2595)$ resonance. Simultaneously, a new resonance in the $I=1$ channel, $\Sigma_c(2800)$, is obtained \cite{,Hofmann:2005sw,Mizutani:2006vq}. 

\subsubsection{SU(6)$\times$HQSS Weinberg-Tomozawa scheme}
\label{su8}

HQSS connects vector and pseudoscalar mesons containing charmed quarks, as all types of spin interactions vanish for infinitely massive quarks.  Chiral symmetry fixes the lowest order interaction between Goldstone bosons and other hadrons  in a model-independent way; this is the WT interaction. Then, it is very appealing to have a predictive model for four flavors including all basic hadrons (pseudoscalar and vector mesons, and $1/2^+$ and $3/2^+$ baryons) which reduces to the WT interaction in the sector where Goldstone bosons are involved and which incorporates HQSS in the sector where charm quarks participate. Indeed, this is a model assumption which is justified in view of the reasonable semiqualitative outcome of the SU(6) extension in the three-flavor sector \cite{Gamermann:2011mq} and in a formal plausibleness on how the SU(4) WT interaction in the charmed pseudoscalar meson-baryon sector comes out in the vector-meson exchange picture. 

The model obeys SU(6) spin-flavor symmetry and also HQSS  \cite{Garcia-Recio:2013gaa}. This is a model extension of the WT SU(3) chiral Lagrangian \cite{GarciaRecio:2008dp,Romanets:2012hm}. The extended SU(6)$\times$HQSS WT meson-baryon interaction is given by
\begin{equation}
V_{ij}^{IJSC}(s)= \frac{D_{ij}^{IJSC}}{4\,f_i f_j}
(2\sqrt{s}-M_i-M_j) \sqrt{\frac{M_i+E_i}{2M_i}}
\sqrt{\frac{M_j+E_j}{2M_j}} 
\,.
\label{eq:vsu8break}
\end{equation}
Again, $i$ ($j$) are the outgoing (incoming) meson-baryon channels. The quantities $M_i$, $E_i$ and $f_i$ stand for 
the baryon mass and energy, in the center-of-mass frame, and the meson decay constant in the $i$ channel, respectively. $D_{ij}^{IJSC}$  are the matrix elements coming
 from the group structure, while the symmetry breaking is now introduced by using physical masses and decay constants. 

We focus once more on the transitions involving $D$ mesons with nucleons. Moreover, the inclusion of $3/2^+$ baryons allows us to study the interaction of $D$ mesons
with $\Delta$. The different channels involved in both cases can be found in Refs.~\cite{GarciaRecio:2008dp}. The Bethe-Salpeter equation is then solved in coupled channels by using the SU(6)$\times$HQSS WT interaction, while the loop function is regularized by means of a subtraction point prescription \cite{Nieves:2001wt}. The $\Lambda_c(2595)$ is generated dynamically together with several resonances in the $C=1$ and $S=0$, some of which can be directly identified with existing  experimental states \cite{GarciaRecio:2008dp,Romanets:2012hm}.

The vacuum $\Delta$-decay width has been considered for the determination of the dynamically-generated resonances and, thus, for the scattering
amplitudes. This effect is introduced in the unitarization procedure through a convolution of the $D$$\Delta$ propagator with the corresponding
spectral function of the $\Delta$ baryon. Only the resonances that lie close to the $D$$\Delta$ channel, as compared to the $\Delta$ width, and that couple strongly to this system
will be affected. Thus, we might expect changes in the $D$$\Delta$ scattering amplitudes around the $D$$\Delta$ threshold.

There are several differences between the SU(6)$\times$HQSS WT scheme and the SU(4) WT model. One of the main differences comes from  the inclusion, on equal footing, of heavy pseudoscalar and vector mesons as  well as baryons with $J^P=1/2^+$ and $J^P=3/2^+$. Another essential difference lies in the transition amplitudes. According to Eq.~(\ref{eq:vsu8break}) of the SU(6)$\times$HQSS WT scheme,  the amplitudes scale with the inverse values of the meson decay constants, whereas in the SU(4) model the decay constants are all fixed to $f=1.15 f_{\pi}$ and the charm-exchange transitions are suppressed with a $\chi_c$ factor. Note, however, that both extensions of the SU(3) WT respect HQSS constraints (see, for example, discussions in Refs.
\cite{Romanets:2012hm,Garcia-Recio:2013gaa,Xiao:2013jla}).

Note that the use of Weinberg-Tomozawa-like structure is an approximation to the full analytical structure of the $D$-meson -- baryon interaction. However, to our knowledge, these are the most up-to-date models used for the interaction
of $D$ mesons with baryons. Only very recent calculations with the J\"ulich exchange model~\cite{Haidenbauer:2007jq,Haidenbauer:2010ch} have dealt with the more complicated analytical structure of the
interaction, including $s$-, $t$- and $u$-channel contributions as well as higher partial waves. However, this latter model
is not consistent with heavy-quark spin symmetry, which is a proper QCD symmetry in the limit of heavy masses, since it does not include the vector-meson baryon channels. We have thus chosen a more simplified interaction that is consistent with this symmetry, the SU(6)xHQSS model, and analyzed the effects with respect to a similar interaction from the SU(4) scheme but where the HQSS constraints are not obeyed.

\section{Results for transport coefficients\label{sec:results}}

In this section we show our results for the drag force and diffusion coefficients of Eqs.~(\ref{eq:Fcoeff},\ref{eq:G0coeff},\ref{eq:G1coeff}) as a function of
temperature and baryonic chemical potential. The required multidimensional integrations are performed using a Monte Carlo routine as explained in detail in
Ref.~\cite{Abreu:2011ic}.  

The main uncertainty in the calculation comes from the existence of an ultraviolet cutoff scale  in the effective theory description of the scattering
amplitudes.  The Monte Carlo routine requires some momentum integrations, whose upper limits are, thus, truncated according to this cutoff in order
 to avoid an uncontrolled behavior of the scattering amplitudes beyond that scale. The error introduced by truncating the integral domain is minimal at low temperatures, as
the average momentum is much lower than the ultraviolet cutoff. However, the error increases as we move to higher temperatures. We have checked, though, that for
temperatures $T=m_{\pi} \sim 140$ MeV ---close to the transition temperature--- the numerical error is negligible. 

Moreover, for our calculations we take $p=100$ MeV for the momentum of the $D$ meson. This value effectively corresponds to the so-called {\it static limit}, where the $D$-meson
momentum is sent to zero. This choice has a threefold goal. First, this is the limit usually taken for the calculation of the transport coefficients. Thus, we will be able to compare our results to previous works. Second,  we can provide a consistency check of our results since only in this limit does one have the property 
\be \label{eq:staticdiffusion} \lim_{p\rightarrow 0} \left[ \Gamma_0(p) - \Gamma_1 (p) \right] = 0 \ , \ee
which results from  Eqs.~(\ref{eq:G0coeff},\ref{eq:G1coeff}). Furthermore, the Einstein relation between the drag force $F$ and diffusion coefficient in momentum space
$\Gamma=\Gamma_0=\Gamma_1$, defined as
\be \label{eq:einstein} F = \frac{\Gamma}{m_DT} \ , \ee
is fulfilled when $p\rightarrow 0$. Finally, in the static limit one has access to the spatial diffusion coefficient, given by~\cite{Abreu:2011ic}
\be \label{eq:dx} D_x= \lim_{p \rightarrow 0} \frac{\Gamma (p)}{m_D^2 F^2 (p)} \ . \ee
The spatial diffusion coefficient $D_x$ appears in Fick's diffusion law and it has been widely studied in the past~\cite{fick}.
The classical behavior of $D_x$ for a nonrelativistic Brownian particle with mass $m_D$ in a bath composed by particles of mass $m_l$ is well known~\cite{smith1989transport,chapman1952mathematical}
\be \label{eq:dxnonrel} D_x \sim \sqrt{ \frac{1}{m_D}+ \frac{1}{m_l}} \ \frac{T^{3/2} }{P \sigma} \ , \ee
where $\sigma$ is the total cross section and $P$ is the pressure
\be \label{eq:pressure} P \sim T^{5/2} m_l^{3/2} e^{\frac{\mu-m_l}{T}} \ , \ee
with $\mu$ being the chemical potential of the light particles in the bath.
Using Eqs.~(\ref{eq:einstein},\ref{eq:dx},\ref{eq:dxnonrel}) one can obtain the nonrelativistic expressions for the drag force and diffusion coefficient
in the static limit $p \rightarrow 0$ when $m_l \ll m_D$:
\ba F &\sim&  P \sigma \sqrt{\frac{m_l}{T}} \frac{1}{m_D} \  ,\\ 
\Gamma &\sim& P \sigma \sqrt{m_l T} \ , \ea
which also serve as a consistency check for our computations.

\subsection{Transport coefficients at $\mu_B=0$}

We start by considering the system at vanishing baryochemical potential, so that the net baryon number is zero. In this case, hadrons will be created after a high-energetic collision, like those at the LHC or RHIC (at its top colliding energies).

\begin{figure}
\centering
\includegraphics[width=0.45\textwidth]{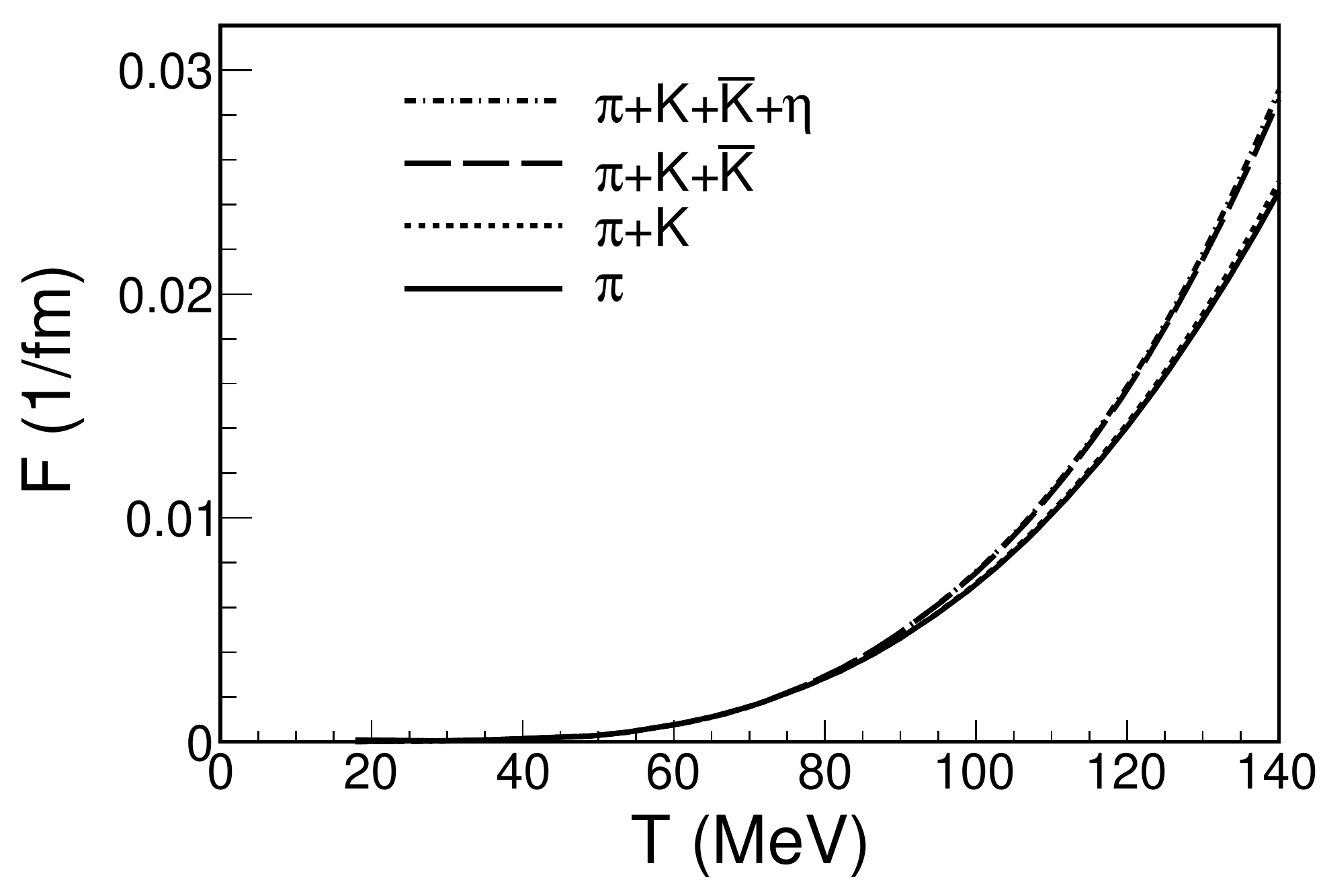}
\caption{\label{fig:Dlight}
$D$-meson -- light-meson contribution to the drag force coefficient $F$  as a 
function of temperature for a $D$-meson momentum of $p=100$ MeV.}
\end{figure}

\begin{figure}
\centering
\includegraphics[width=0.45\textwidth]{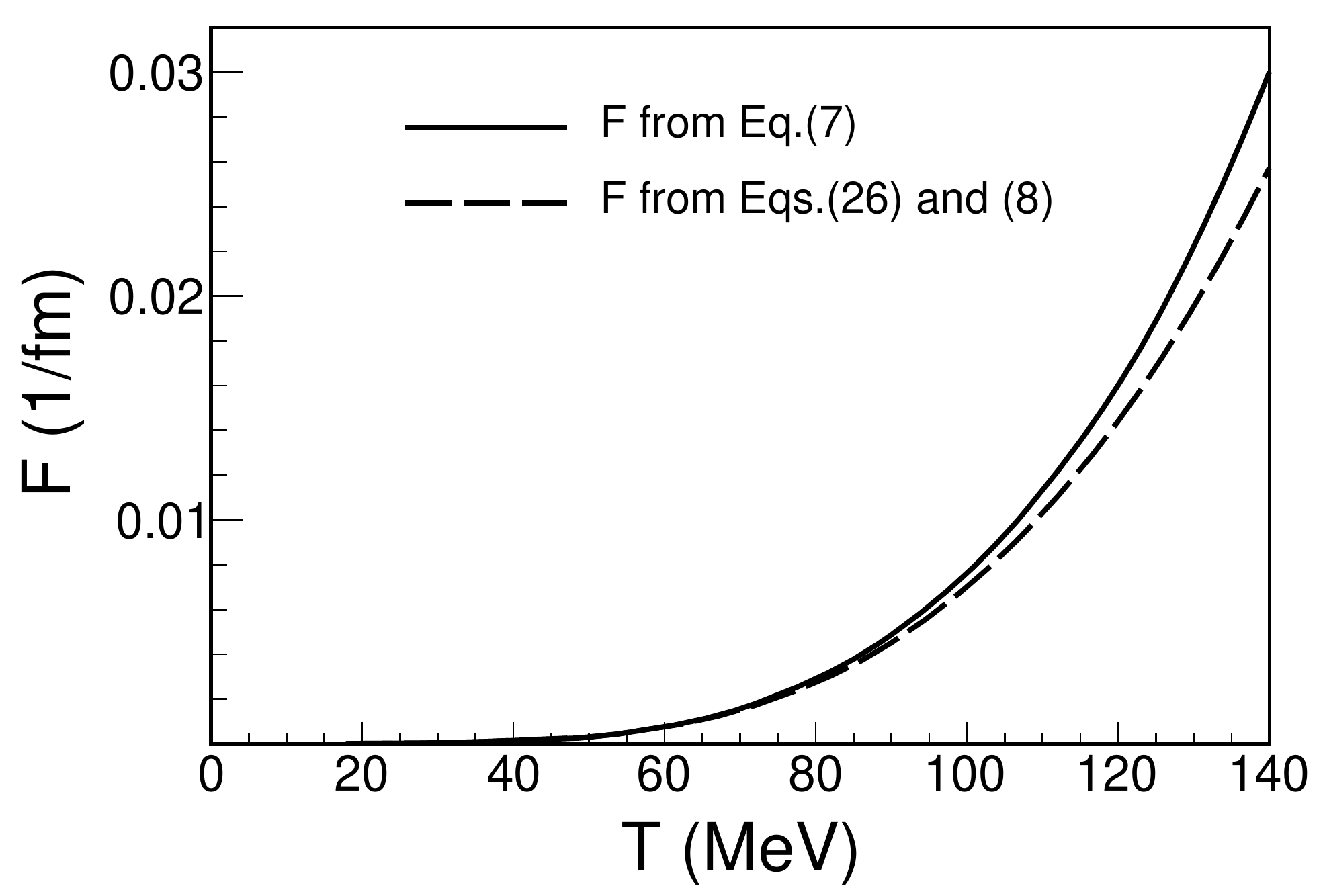}
\caption{\label{fig:einstein}
$D$-meson drag force when all particle species are introduced. We compare the result for this coefficient using Eq.~(\ref{eq:Fcoeff}) and 
obtained from $\Gamma_0$ in Eq.~(\ref{eq:G0coeff}) using the Einstein relation Eq.~(\ref{eq:einstein}).}
\end{figure}

For highly energetic collisions we expect that light mesonic species will mainly populate the thermal bath.  Thus, we first consider matter made of pseudo-Goldstone
bosons, such as pions, kaons, antikaons and $\eta$ mesons. In Fig.~\ref{fig:Dlight} we plot the drag coefficient $F$ as a function of temperature up to $T=140$ MeV ---close
to the transition temperature--- for a $D$ meson with a momentum of $p=100$ MeV when the different Goldstone bosons are consecutively added. We observe that the dominant
contribution is the one associated with the $D\pi$ interaction.  The contribution to the drag force coming from $D$ mesons interacting with pions was already studied
in Ref.~\cite{Abreu:2011ic}. In this previous work, the higher-temperature values for the drag coefficient slightly differ from the ones presented here because, in
that reference, the Einstein relation was used to obtain the drag force in the static limit while we now perform the computation of the $F$ coefficient
directly from its definition in Eq.~(\ref{eq:Fcoeff}), cf. Fig.~\ref{fig:einstein}. However, we obtain identical results for the two diffusion coefficients, $\Gamma_0$ and $\Gamma_1$
as in Ref.~\cite{Abreu:2011ic}.

\begin{figure}
 \centering
\includegraphics[width=0.45\textwidth]{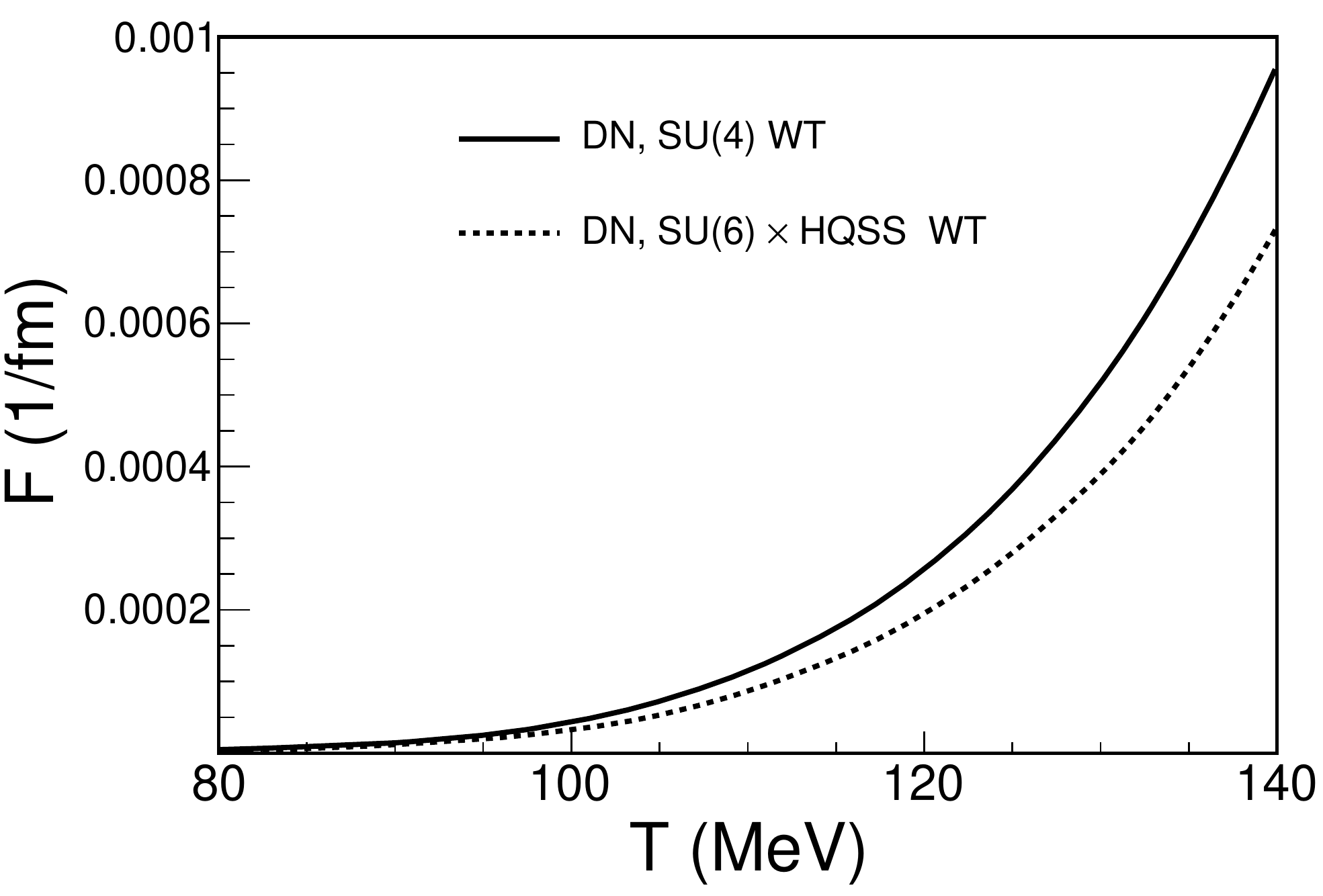}
 \caption{\label{fig:DNcomp}
 $DN$ contribution to the drag force $F$ as a 
 function of temperature for a $D$-meson momentum of $p=100$ MeV. The unitarized scattering amplitudes are taken from the SU(4) and SU(6)$\times$HQSS WT Lagrangians.
}
\end{figure}

\begin{figure}
\centering
\includegraphics[width=0.9\textwidth]{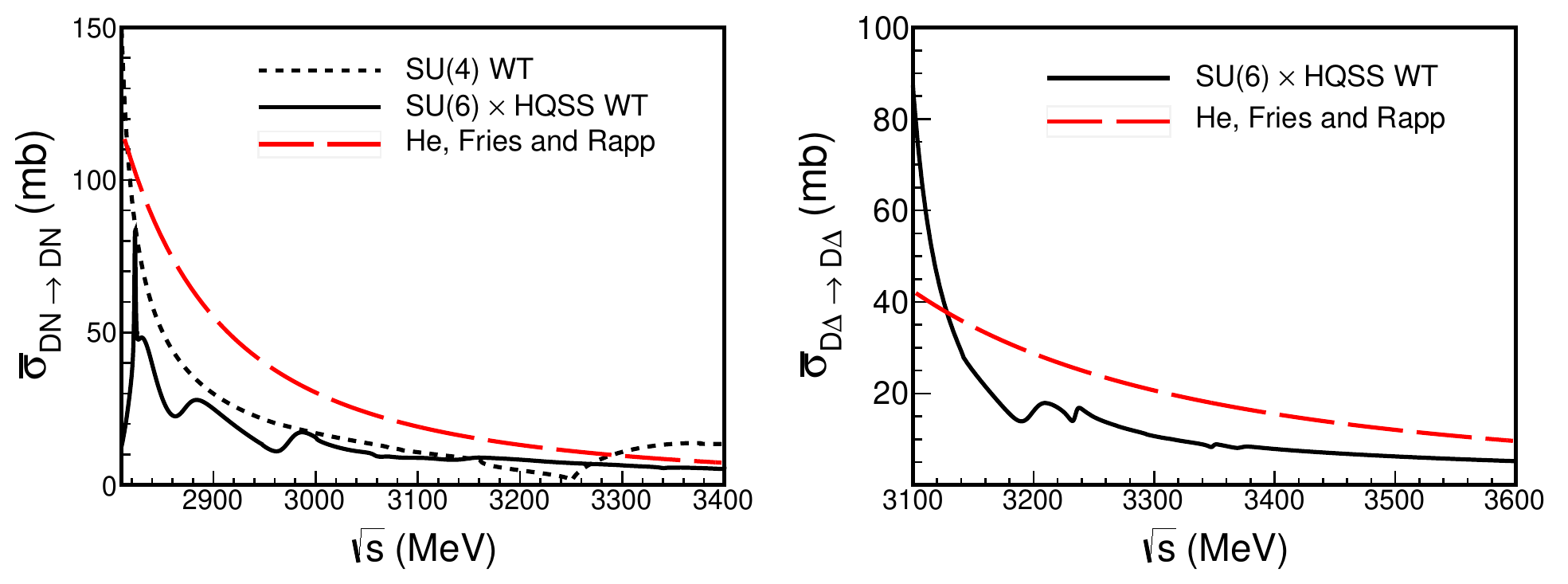}
\caption{\label{fig:crosssections}
Left panel: Isospin averaged $DN \rightarrow DN$ cross section for the SU(4) and SU(6) $\times$ HQSS models. The $DN$ elastic cross section for the SU(4) model is clearly bigger than the one for the SU(6) $\times$ HQSS scheme.
This effect will result in a larger drag coefficient for the former case. Right panel: Isospin averaged cross section for the $D\Delta \rightarrow D\Delta$ scattering in the SU(6) $\times$ HQSS model. For comparison, we also include
the cross sections used in Ref.~\cite{He:2011yi} from a Breit-Wigner parametrization of the resonances.
}
\end{figure}

We now consider the effect on the drag force of the presence of nucleons in the thermal bath. For that purpose, we make use of the two models for the interaction of
$D$ mesons with baryons discussed in Sec.~\ref{dn}. Both models respect HQSS, but in the SU(4) WT scheme only $1/2^+$  baryon are taken into account while for the
SU(6) $\times$ HQSS model $1/2^+$ and $3/2^+$  baryons are also present.  In Fig.~\ref{fig:DNcomp} we show the $DN$ contribution to the transport coefficients in
both schemes.  We observe that the results for the drag force within the SU(4) model are systematically bigger than for the SU(6) $\times$ HQSS case. This can be understood
from Fig.~\ref{fig:crosssections}, as the isospin-averaged cross section for the elastic $DN$ interaction as a function of the energy in the SU(4) scheme is larger than
for the SU(6) $\times$ HQSS model. For comparison, we also include here the results of Ref.~\cite{He:2011yi}, where the scattering amplitudes of the fitted resonances give 
cross sections quite similar to ours; thus, the simple Breit-Wigner parametrization provides a reasonable description of the $DN-DN$ and $D\Delta-D\Delta$ interactions.
 It is interesting to note that the values for the drag force when only considering nucleons in the thermal bath for all temperatures
up to $T\sim 140$ MeV are much smaller, by an order of magnitude, than when considering a bath populated by light mesons. For $\mu_B=0$ and $T \sim 140$ MeV, the
nucleon density with respect to saturation density, $n_0=0.16$ fm$^{-3}$, is $n_N/n_0 \simeq 1.5$ $\times$ $10^{-2}$. The low population of nucleons is responsible for the small contribution to the drag coefficient.

\begin{figure}
\centering
\includegraphics[width=0.45\textwidth]{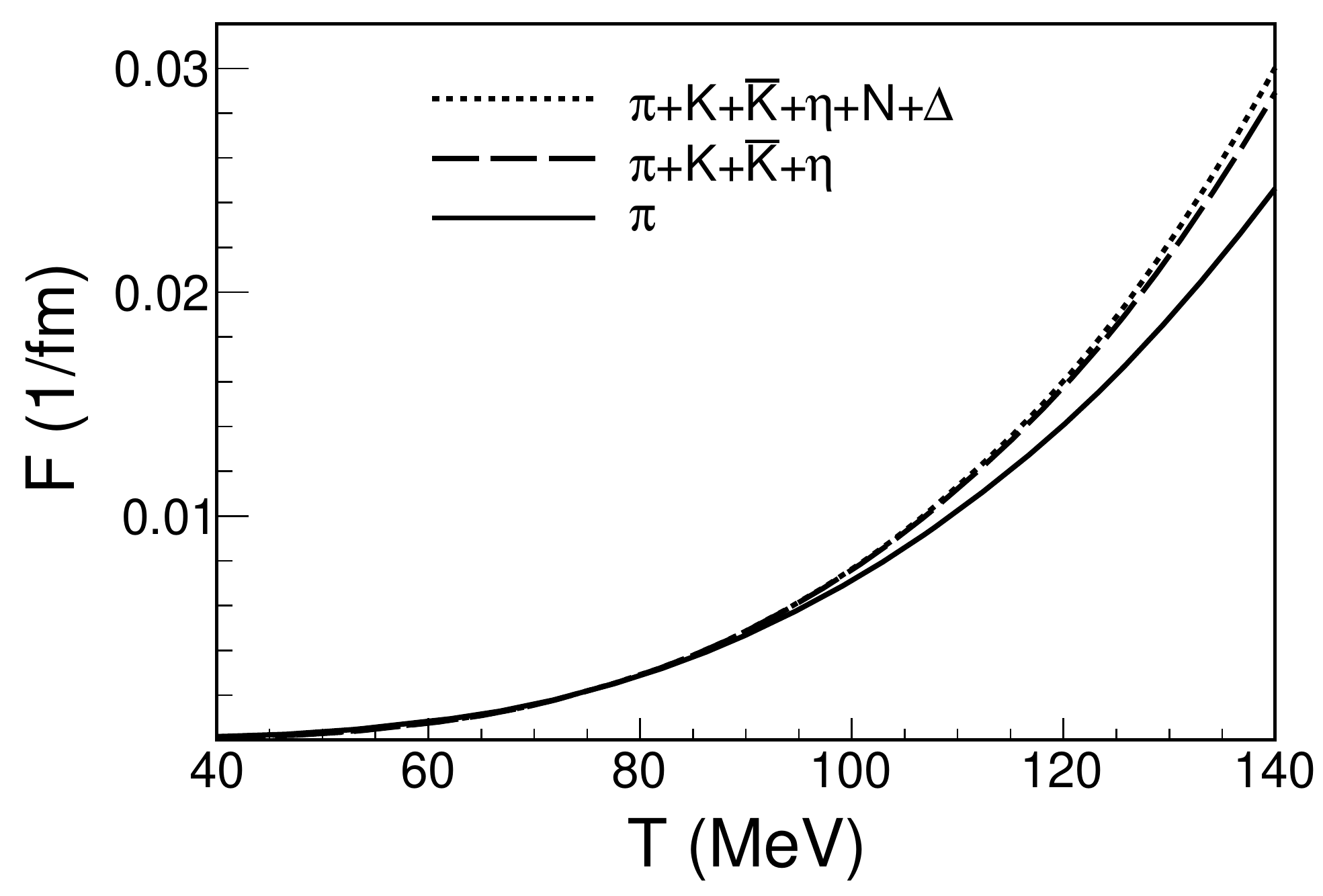}
\includegraphics[width=0.45\textwidth]{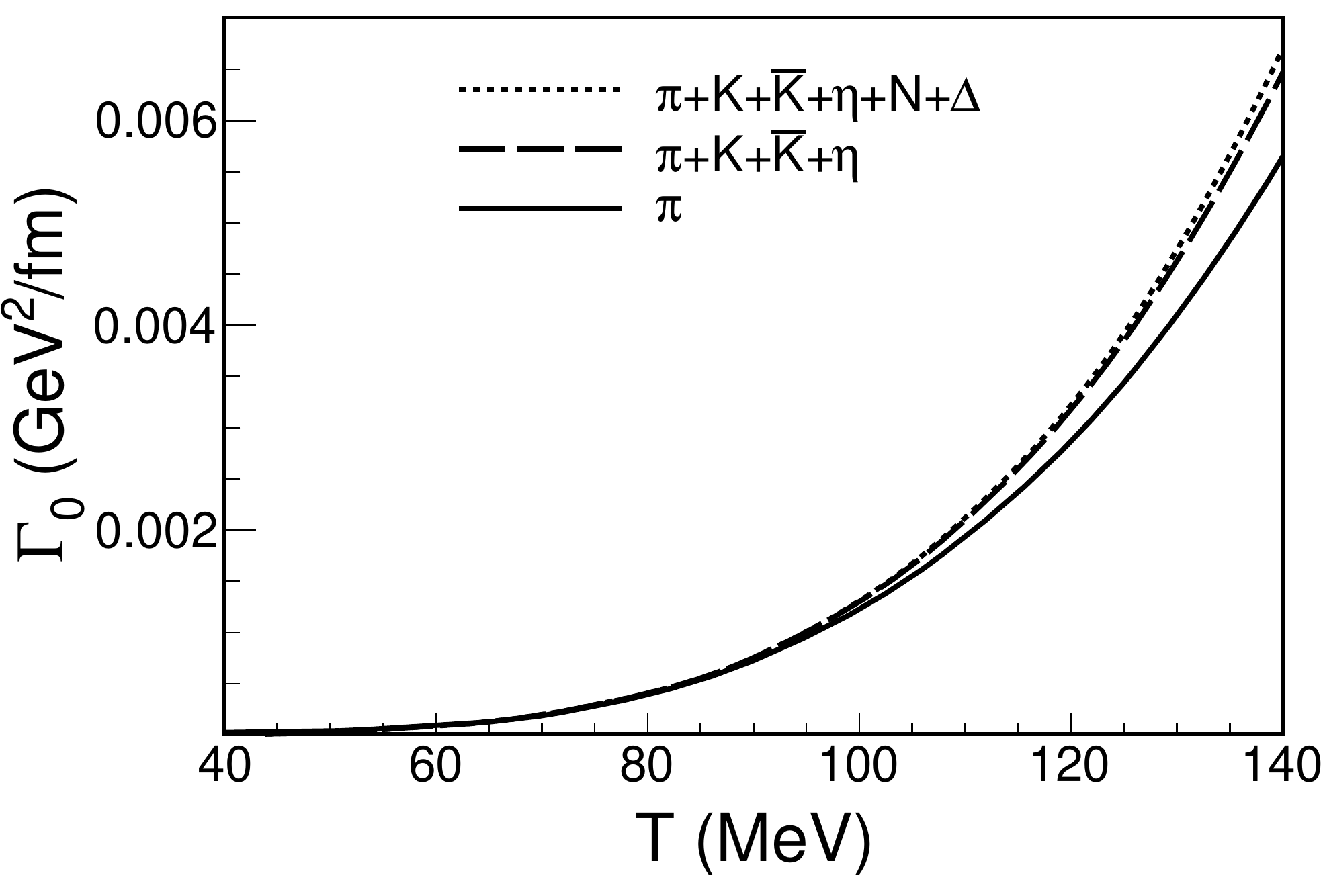}
\includegraphics[width=0.45\textwidth]{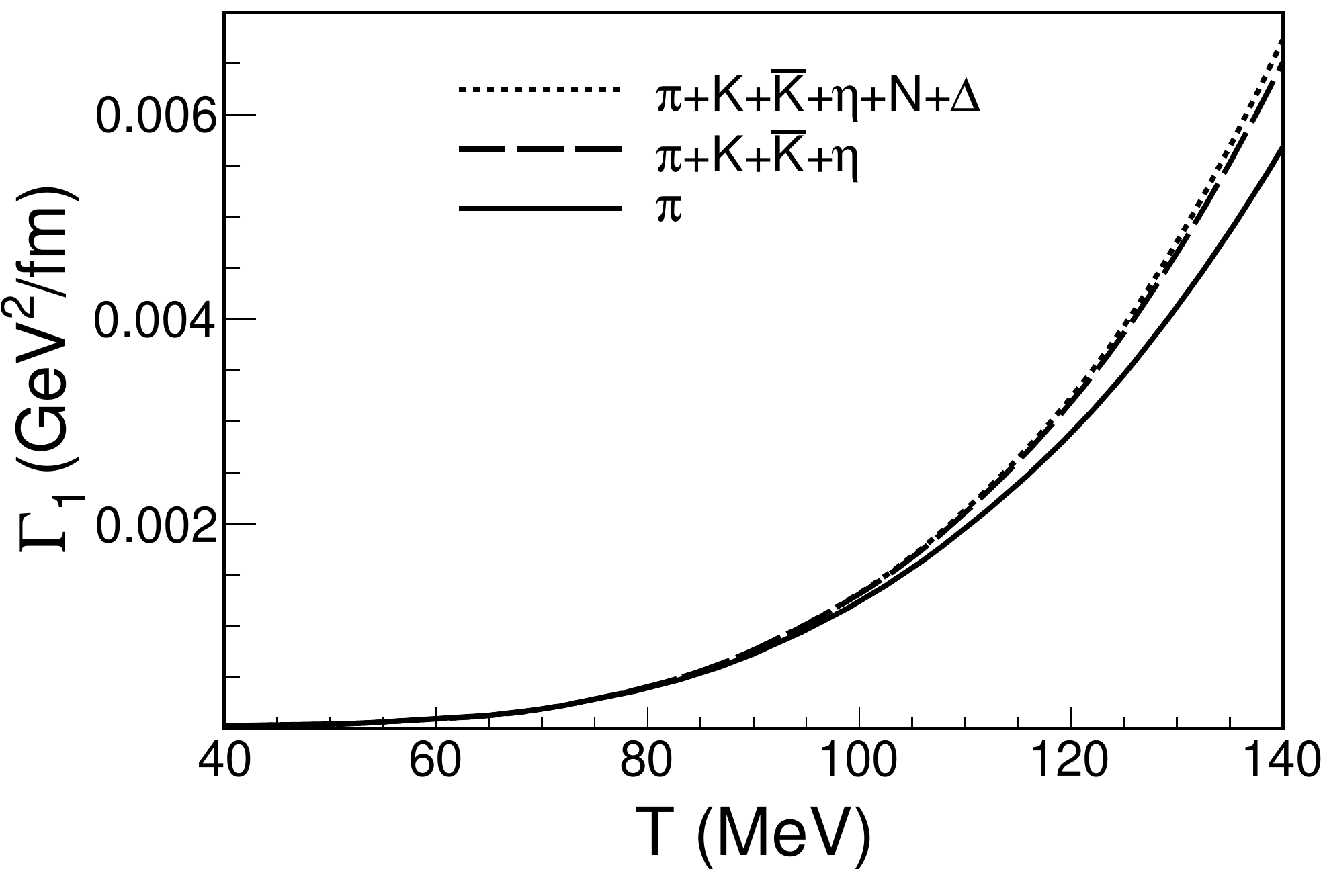}
\caption{\label{fig:finalplot}
Drag force and diffusion coefficients for $D$ mesons when all particle species are introduced. }
\end{figure}

In Fig.~\ref{fig:finalplot} we present the drag and diffusion coefficients at $\mu_B=0$ as function of the temperature for a $D$ meson in a medium of light mesons
together with nucleons and $\Delta$'s, which are the lightest baryons. We make use of the SU(6) $\times$ HQSS model to obtain the interaction of $D$ mesons with
nucleons as well as with $3/2^+$ $\Delta$ baryons. As indicated before, the main contribution  to the transport coefficients comes from the pion gas, as it is the
most populated species in the medium. The other Goldstone bosons give a sizable contribution only for $T \gtrsim 100$ MeV while $N$ and $\Delta$ do not contribute
substantially to the transport coefficients.

Comparing the two plots for the diffusion coefficients, we notice that $\Gamma_0=\Gamma_1$ of Eq.~(\ref{eq:staticdiffusion}) is well satisfied in the range of temperatures studied. Moreover, in Fig.~\ref{fig:einstein} we show the drag coefficient calculated using the Einstein relation of Eq.~(\ref{eq:einstein}), as well as by means of its definition from Eq.~(\ref{eq:Fcoeff}). We observe that the Einstein relation is well satisfied up to $T\sim 100$ MeV, presenting some small violations at high temperatures due to the cutoff effects in the numerical computation of the transport coefficients.

\begin{figure}
\centering
\includegraphics[width=0.45\textwidth]{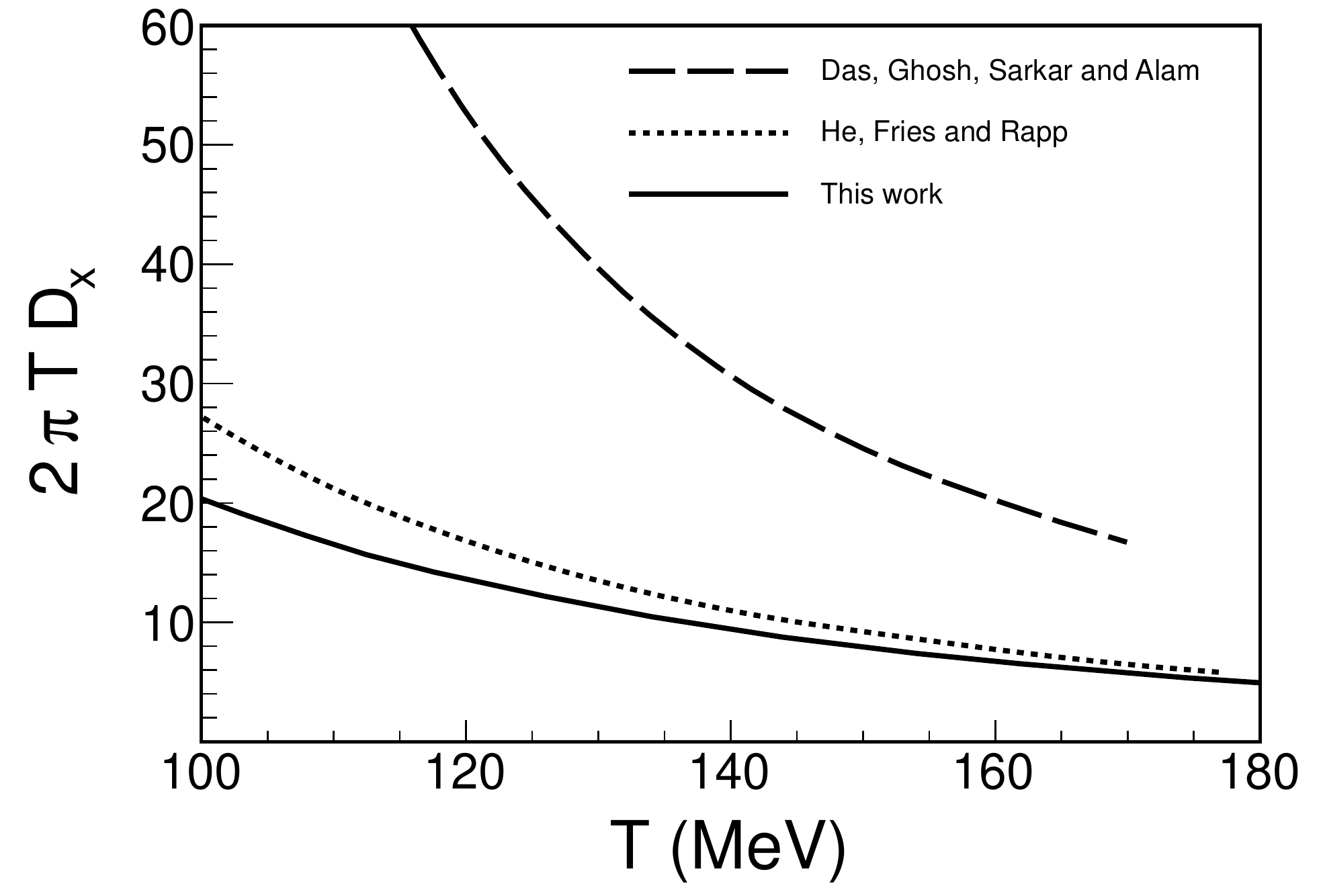}
\caption{\label{fig:comparison}
Spatial diffusion coefficient for $D$ mesons multiplied by $2 \pi T$. Solid line: our work. Dotted line: results from Ref.~\cite{He:2011yi}. Dashed line:
results from Ref.~\cite{Das:2011vba}.}
\end{figure}

Finally, in Fig.~\ref{fig:comparison} we present our results for the spatial diffusion coefficient at 100 MeV $\lesssim T \lesssim$ 180 MeV in order to compare with the
 outcome of Refs.~\cite{He:2011yi} and~\cite{Ghosh:2011bw}. We observe a good agreement with the results of Ref.~\cite{He:2011yi} and substantial differences between
 our outcome and the one from Ref.~\cite{Ghosh:2011bw}. In Ref.~\cite{He:2011yi}  the interaction of $D$ mesons with pions was approximated by $D^*$ resonances, while
 scattering off other hadrons, such as $K$, $\eta$, $\rho$, $K^*$, $N$, $\Delta$, $\bar N$ and $\bar \Delta$, was evaluated from different microscopic models, parametrizing
 certain resonant states and, thus, mimicking the unitarized scattering amplitudes.  As indicated in that work, the scattering of $D$ mesons with vector mesons
  and antibaryons  as well as the inclusion of higher-mass resonant states, does not play an important role in the determination of the drag coefficient for the range of
 temperatures studied. For example, there is not a known resonant state that can be generated dynamically due to $D\bar{N}$ scattering  and the $D\bar{N}$ contribution to the transport
is therefore suppressed (by a factor of 3) with respect to the $DN$ channel, whereas the $D\bar{\Delta}$ relaxation rate is about half of the $D\Delta$ one. Thus, a good agreement is obtained
 between the two approaches. On the contrary, the use of NNLO scattering lengths for the scattering of $D$ mesons with
 mesons and baryons accounts for the very different drag and diffusion coefficients and, hence, the spatial diffusion coefficient of Ref.~\cite{Das:2011vba}. The
 temperature dependence of the spatial diffusion coefficient can be inferred in the static limit making use of the nonrelativistic expressions of Eqs.~(\ref{eq:dxnonrel},\ref{eq:pressure}). In this case
\be T D_x \sim e^{-\frac{\mu_B-m_l}{T}} \ , \ee
showing that the combination $TD_x$ must decrease with temperature for constant chemical potential (in particular for $\mu_B=0$) as seen in our results.

\subsection{Transport coefficients for isentropic trajectories}

\begin{figure}
\centering
\includegraphics[width=0.45\textwidth]{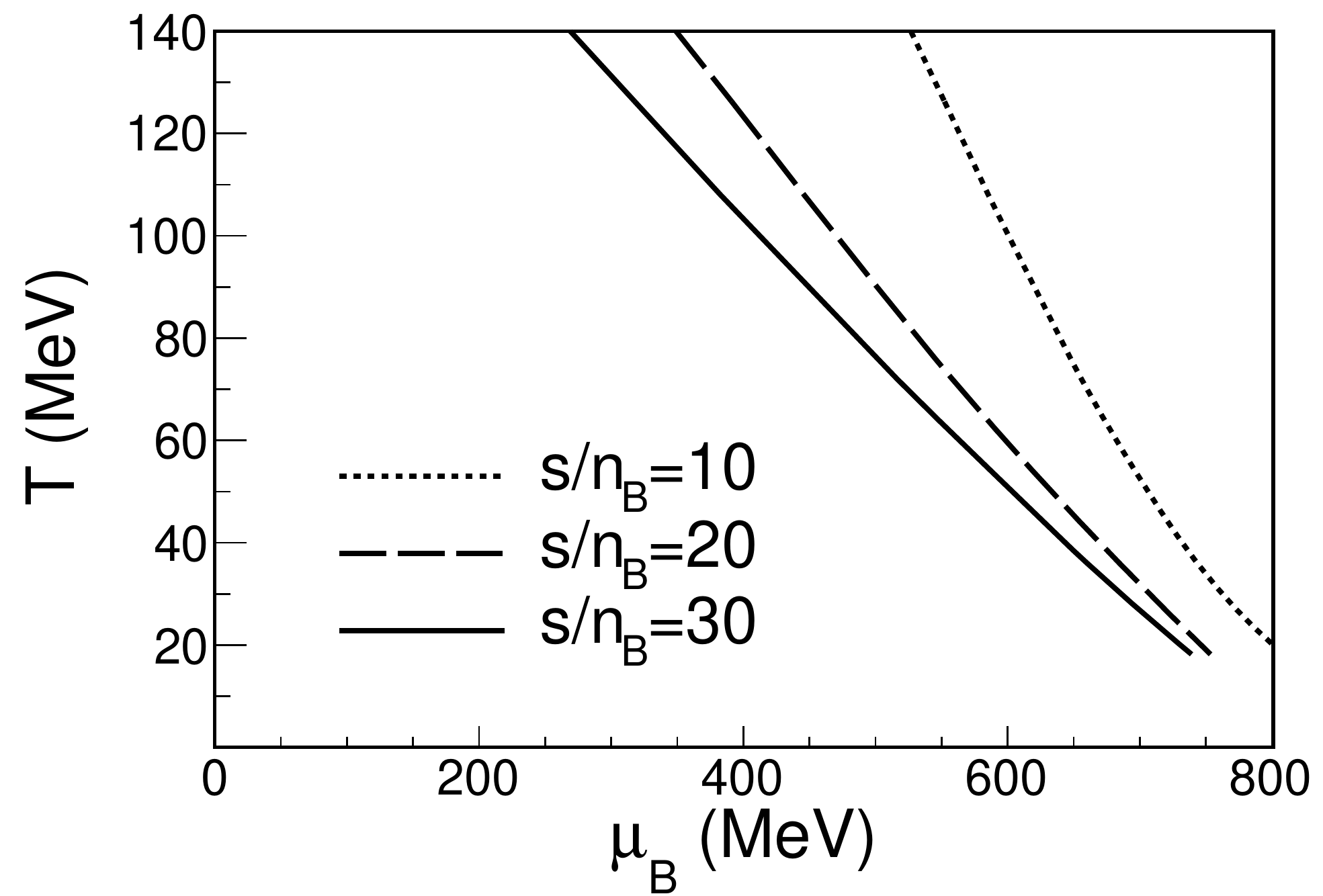}
\caption{\label{fig:adiab} Three characteristic trajectories for hadronic matter in the QCD phase diagram. The entropy per baryon ($s/n_B$) is constant along those trajectories and
its value is fixed for three characteristic FAIR energies in the range $\sqrt{s}=5-40$ $A$GeV.}
\end{figure}

Heavy-ion collisions at the future facilities of FAIR or NICA will run at lower beam energies than the LHC. Collisions at FAIR  will be produced with a beam energy such
that the hadronic trajectories will cross the phase diagram through the finite baryochemical potential region. However, the baryonic density will not be constant during
the evolution. The simplest realization of physical trajectories for FAIR energies consists of nearly constant entropy per baryon trajectories, $s/n_B={\rm constant}$.
Although the dissipative phenomena (transport coefficients) always present entropy production in the system, an isentropic trajectory is a good approximation for the
physical evolution of the fireball. We will assume constant $s/n_B$ and extract the associated baryochemical potential, $\mu_B=\mu_B(T)$, according to this criterion.
In particular, for FAIR physics, the beam energy runs from $\sqrt{s}=5-40$ $A$GeV, which approximately corresponds to $s/n_B=10-30$, with some dependence on the thermal model
used~\cite{Bravina:2008ra}. We choose three characteristic values of $s/n_B=10,20$ and $30$ and show the expected isentropic trajectories in Fig.~\ref{fig:adiab}.

\begin{figure}
\centering
\includegraphics[width=0.45\textwidth]{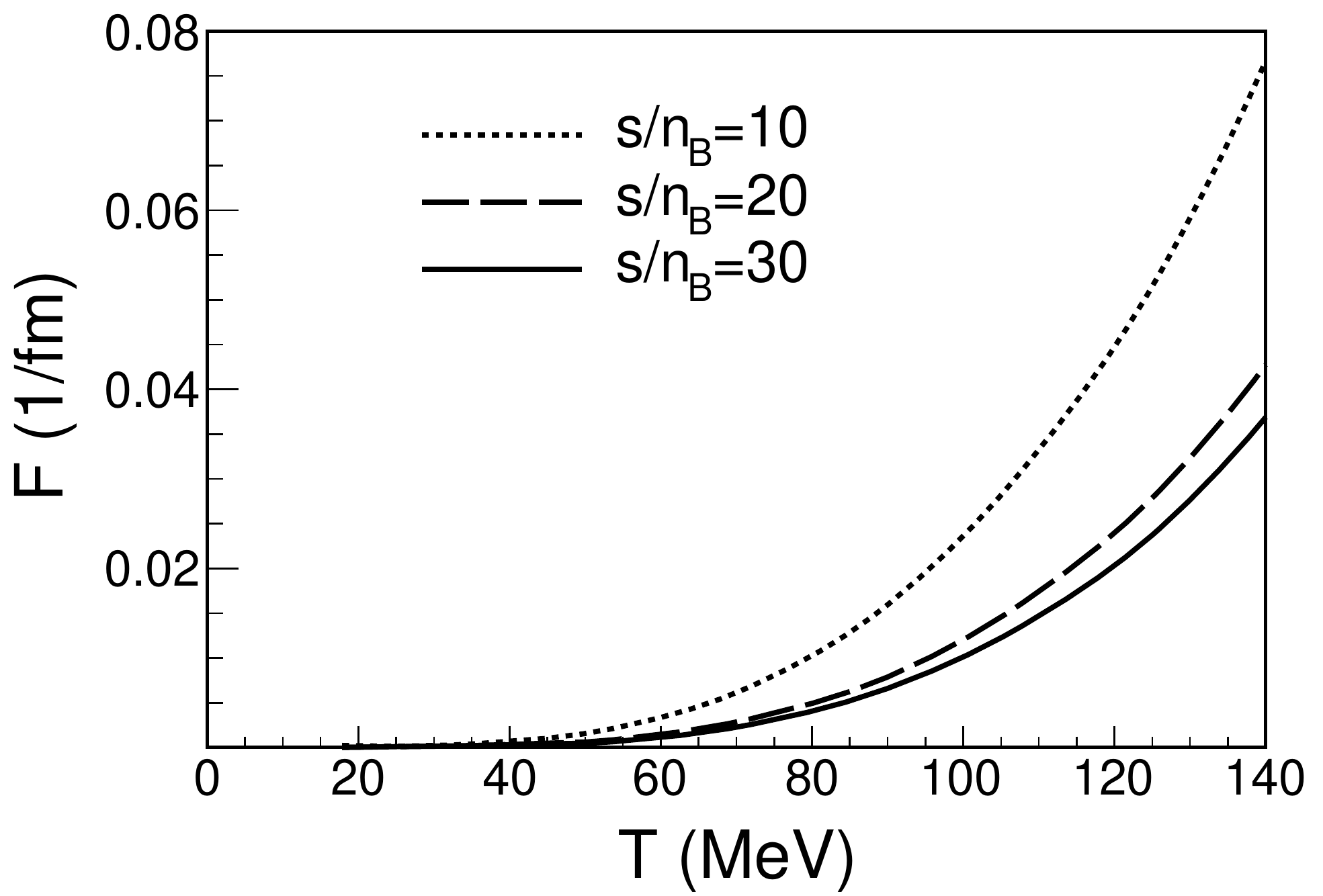}
\includegraphics[width=0.45\textwidth]{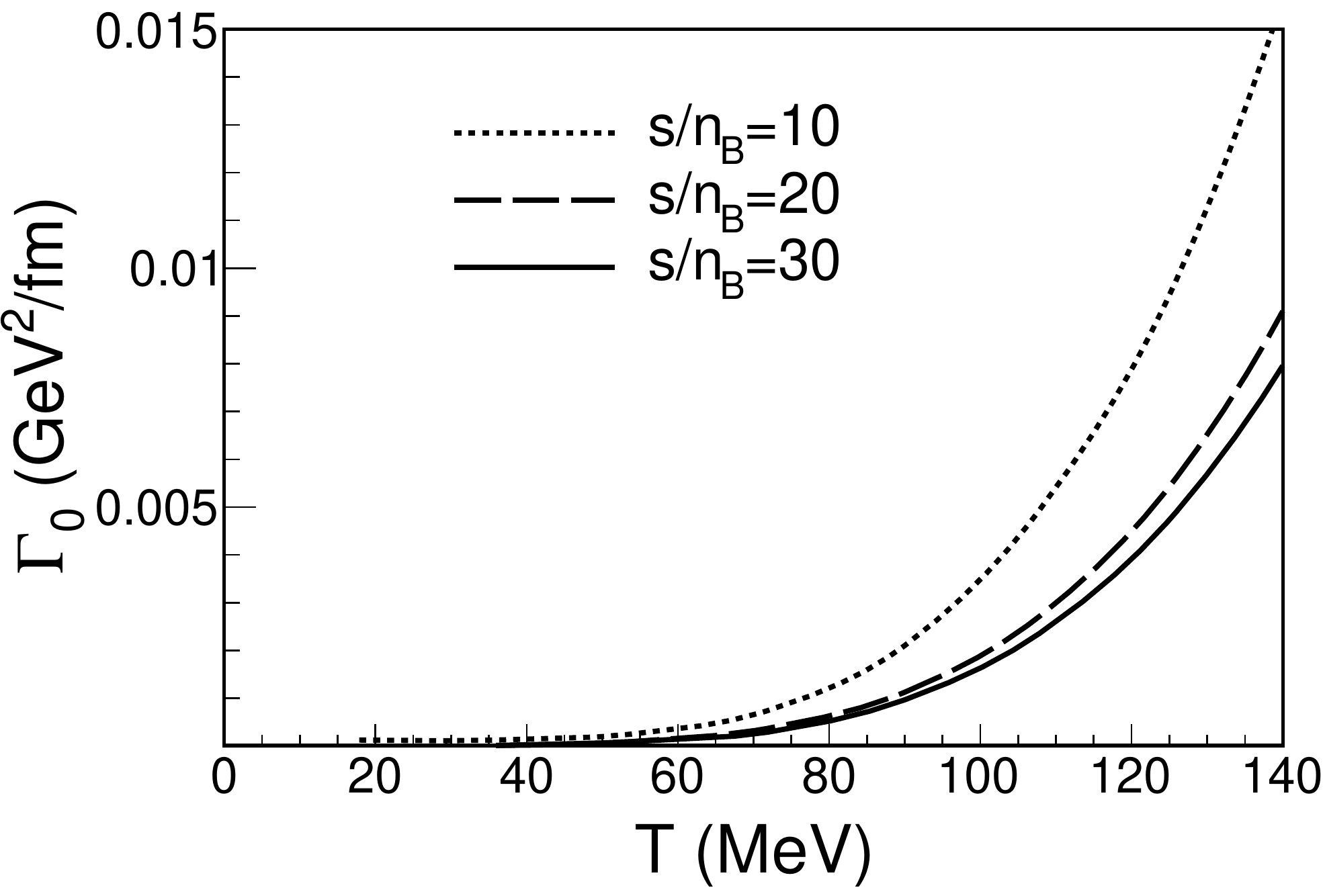}
\includegraphics[width=0.45\textwidth]{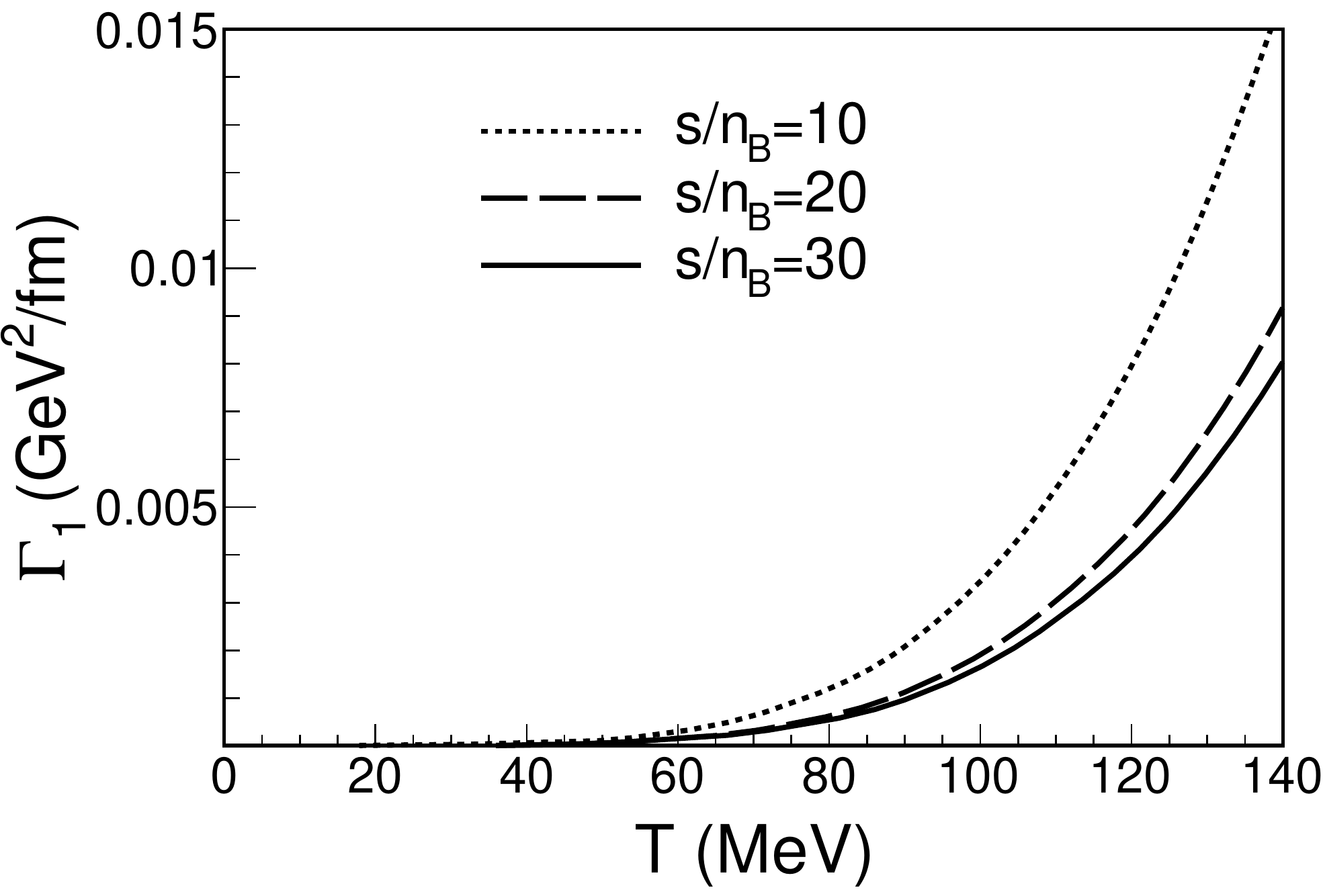}
\caption{\label{fig:adiabcoeff} Drag force and diffusion coefficients of a $D$ meson in a hadronic gas for the three isentropics  detailed in Fig.~\ref{fig:adiab}.}
\end{figure}

For the specific trajectories of Fig.~\ref{fig:adiab}, we show in Fig.~\ref{fig:adiabcoeff} the transport coefficients of a $D$ meson embedded in a thermal bath of light
mesons, nucleons and $\Delta$'s . Note that the transport coefficients strongly depend on the isentropic trajectory. In addition, by comparing the transport coefficients
of a $D$ meson embedded in a bath of light mesons  of Fig.~\ref{fig:finalplot} with our results in the thermal bath for an isentropic trajectory (and given that the 
contribution of light mesons only depends on the temperature) we observe that nucleons and $\Delta$'s contribute significantly to the transport coefficients for finite 
baryochemical potential. This is mainly due to the fact that the density of nucleons has increased significantly, the exact value depending on the temperature and 
baryochemical potential at each given isentropic trajectory. 

It is worth mentioning that the $D$-baryon contribution to $F$ and $\Gamma$ follows a very simple scaling with the fugacity:
\be F (T,\mu_B) \simeq F (T,\mu_B=0) \ e^{\mu_B/T} \ , \ee
\be \Gamma (T,\mu_B) \simeq \Gamma (T,\mu_B=0) \ e^{\mu_B/T} \ . \ee
These relations can be obtained from Eqs.~(\ref{eq:Fcoeff},\ref{eq:G0coeff},\ref{eq:G1coeff}) assuming that the classical statistics can be used instead of the Fermi-Dirac distribution.
Notice that this contribution should be added to the mesonic contribution which does not depend on $\mu_B$~\footnote{The mesonic contribution to the transport
coefficents depend on $\mu_B$ when considering in-medium effects on the scattering amplitudes. In this work the scattering amplitudes are calculated in vacuum.
The study of the nuclear medium effects on the unitarized amplitudes and transport coefficients is left for future work.}.
We can thus conclude that the presence of baryons enhances the stopping of the $D$ mesons in the finite baryochemical region of the QCD phase diagram.

\begin{figure}
\centering
\includegraphics[width=0.45\textwidth]{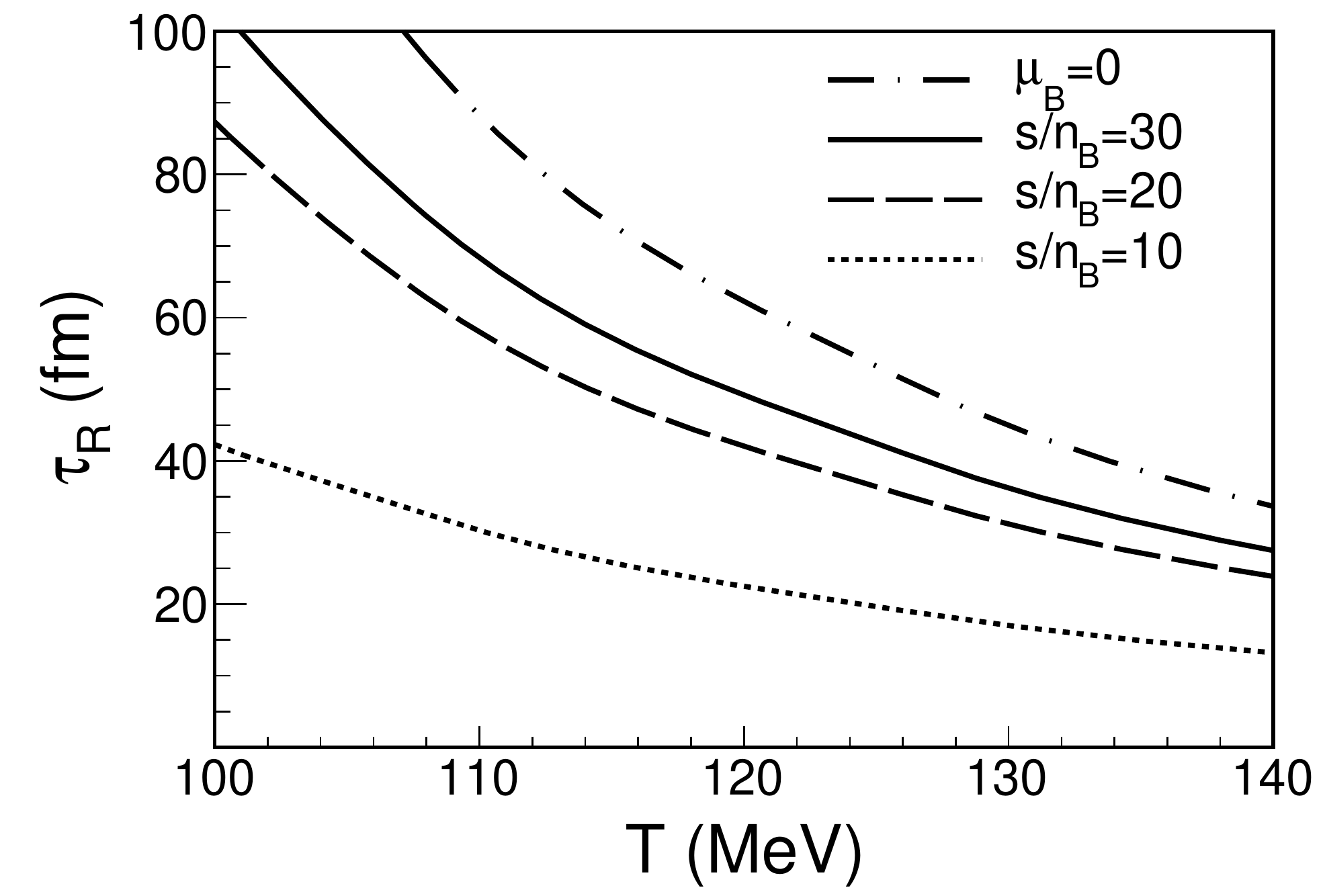}
\caption{\label{fig:relaxation} $D$-meson relaxation time as a function of temperature for the $\mu_B=0$ and the three isentropic trajectories.}
\end{figure}

\begin{table}
\begin{ruledtabular}
\begin{tabular}{cccc}
Trajectory & $\mu_B$ (MeV) & $n/n_0$ & $\tau_R $(fm) \\
\hline
$\mu_B=0$ & 0 & 0 & 34.5 \\
$s/n_B=30$ & 286 & 0.11 & 28.3 \\
$s/n_B=20$ & 361 & 0.20 & 24.3 \\
$s/n_B=10$ & 536 & 0.68 & 13.4 
\end{tabular}
\end{ruledtabular}
\caption{\label{tab}Baryochemical potential, net nuclear density and relaxation time at $T=140$ MeV for the trajectories shown in Fig.~\ref{fig:adiab}. The normal nuclear density is denoted by $n_0=0.16$ fm$^{-3}$.}
\end{table}

In order to have a quantitative measure of the stopping and, hence, the degree of thermalization of $D$ mesons in the medium, we can calculate the typical time
associated to the relaxation of $D$ mesons in a thermal bath. 

In the Langevin approach, $F$ appears as a deterministic drag force while $\Gamma$ measures the strength of the stochastic force $\xi(t)$:
\be \frac{d\mathbf{p}}{dt}=-F \mathbf{p} + \mathbf{\xi}(t) \ . \ee
For constant $F$ one can formally solve the equation 
\be \mathbf{p} (t)=\mathbf{p}_0 \ e^{-F t} + e^{-Ft} \int_{0}^t d\tau e^{F\tau} \xi(\tau)  \ee
and consider the solution for the averaged momentum:
\be \langle\mathbf{p} (t) \rangle=  \mathbf{p}_0 \ e^{-Ft} \ . \ee
Therefore, one can define a relaxation time $\tau_R$ as the inverse of the drag force coefficient, $\tau_R=1/F$.

In Fig.~\ref{fig:relaxation} we show the relaxation time $\tau_R$ as a function of temperature for the three characteristic isentropic trajectories together with the $\mu_B=0$ case.
The values for the relaxation times at $T = 140$ MeV for the $s/n_B=10, 20, 30$ and $\mu_B=0$ trajectories are also presented in Table~\ref{tab}, where we also
display the associated baryochemical potential $\mu_B$ and the corresponding net baryonic density $n/n_0$.

We observe that for $\mu_B=0$ the relaxation time at $T=140$ MeV reads $\tau_R = 33$ fm. In fact, when only pions are considered in the thermal bath, we obtain 
$\tau_R=40$ fm, as in Ref.~\cite{Abreu:2011ic}. The more species the thermal bath has, the more collisions the $D$ meson suffers and, the shorter the equilibration time is.
Moreover, we see that the relaxation time grows with decreasing temperature following the inverse behavior of the drag force coefficient. It is also interesting to see from Fig.~\ref{fig:relaxation} that the $\mu_B=0$ case can be recovered for $s/n_B \rightarrow \infty$. The bigger the value of $s/n_B$, the smaller the baryochemical potential is and, then, the bigger the value of the thermal relaxation time becomes, as indicated in Table~\ref{tab}.

This relaxation time  needs to be compared to the duration of the fireball expansion. The freeze-out time for a central Pb-Pb collision at the LHC (obtained by
performing pionic Hanbury-Brown-Twiss interferometry) is around 10--11 fm~\cite{Aamodt:2011mr}. Therefore, by looking at the $\mu_B=0$ trajectory obtained for LHC
energies, we can conclude that the $D$ mesons are only slightly thermalized in these collisions (note that the time scale from the hadronization until the freeze-out is even shorter). We also observe
that, by comparing the trajectories at $\mu_B=0$ for the LHC and $s/n_B=10$ for the lowest FAIR energies,  the relaxation time is reduced by $60 \%$ due to
medium effects.

\section{Minimum of $2\pi T D_x$ at the phase transition\label{sec:dif}}

In this section we analyze the behavior of the spatial diffusion coefficient $D_x$ around the deconfinement phase transition. This study is motivated by the fact that
previous works on viscous coefficients, such as the shear and bulk viscosities, show an extremum close to the phase-transition temperature. The shear viscosity, when
normalized to the entropy density, seems to present an absolute minimum at the QCD phase transition~\cite{Csernai:2006zz,Dobado:2008vt}.  A plausible argument to this
claim ---at least for vanishing baryochemical potential--- is given by the results in the asymptotic limits, i.e. the hadronic gas at $T<T_c$~\cite{Dobado:2008vt} (where $\eta/s$ rapidly
decreases when temperature is increased) and the perturbative quark-gluon plasma at $T>T_c$~\cite{Arnold:2003zc} (where $\eta/s$ slowly increases with temperature). 
This minimum is also supported by other examples of different physical systems like the liquid-gas phase transition~\cite{Csernai:2006zz,Dobado:2008vt} or cold Fermi atoms close to
unitarity~\cite{Schafer:2009dj}; and models like the linear sigma model~\cite{Chakraborty:2010fr,Dobado:2009ek} or the Nambu-Jona-Lasinio model~\cite{Sasaki:2008um}. 
Also a maximum of the bulk viscosity over entropy density $\zeta/s$ is found close to the critical temperature of several models~\cite{Chakraborty:2010fr,Khvorostukhin:2010aj,Dobado:2012zf}. 
The breaking of conformal invariance (to which the bulk viscosity is highly sensitive) is maximal at the critical temperature giving a peak of $\zeta/s$ at $T_c$. There
exist some indications that a similar maximum of $\zeta/s$ may happen to QCD~\cite{Kharzeev:2007wb,Karsch:2007jc}.

\begin{figure}
\centering
\includegraphics[width=0.45\textwidth]{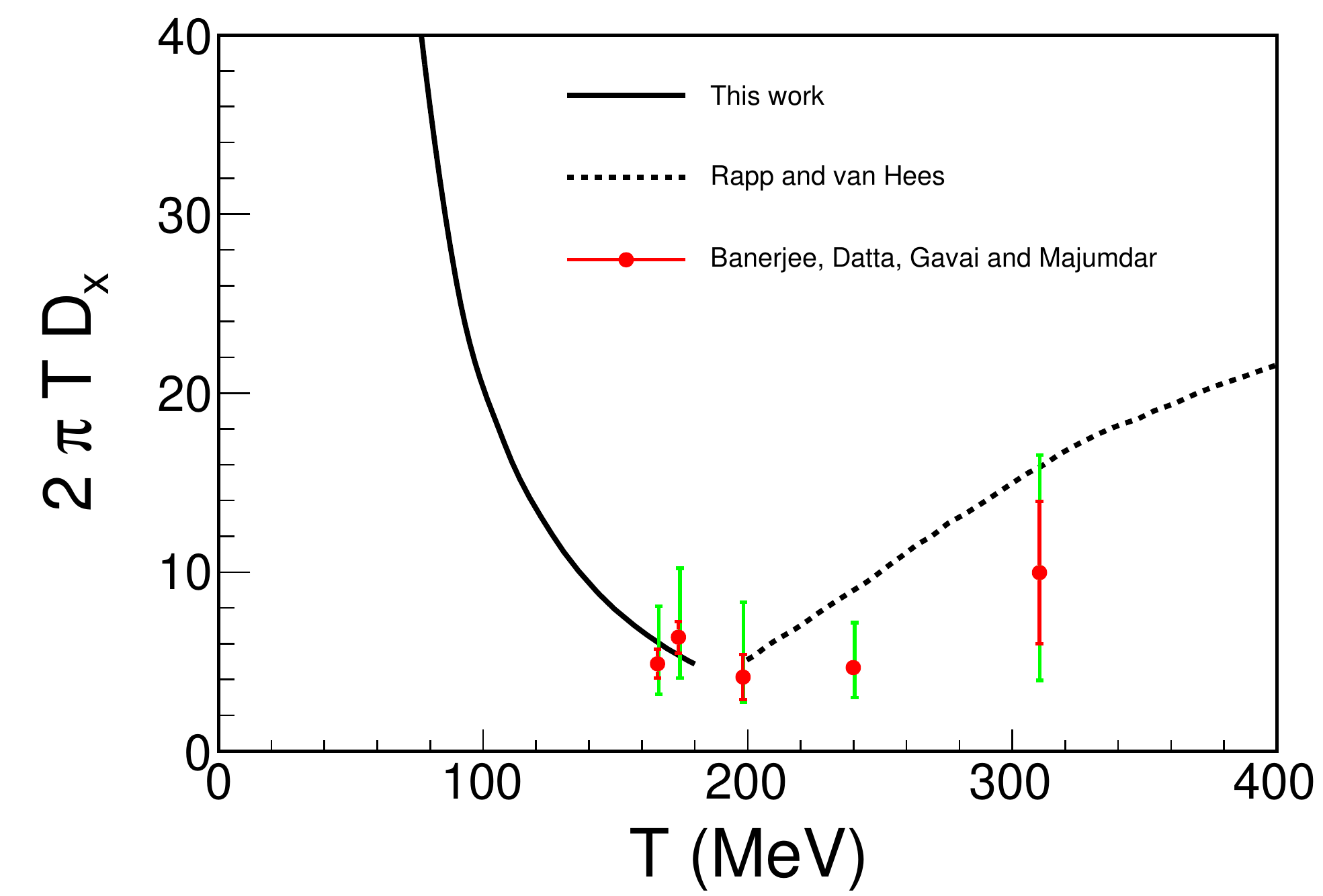}
\caption{\label{fig:transition} Spatial diffusion coefficient for $D$ mesons multiplied by $2\pi T$ in the hadronic phase below $T_c=160$ MeV and the
QGP phase above $T_c$. For the QGP side, we show the results given in Refs.~\cite{Rapp:2008qc} and \cite{Banerjee:2011ra}.}
\end{figure}

 In an analogous way, we can study the behavior of the adimensional number $2\pi T D_x$ near the deconfinement temperature in order to check whether there is also
an extremum. In Fig.~\ref{fig:transition} we plot $2\pi T D_x$ as a function of temperature around the crossover temperature at $\mu_B=0$. For the hadronic side we plot our
results for vanishing chemical potential. For the quark-gluon plasma phase we show the result of Ref.~\cite{Rapp:2008qc}, where a lattice-QCD-based potential for
the light quark--$c$ quark scattering is used to construct the $T$ matrix. The $T$ matrix contains the medium effects and it is supplemented by a perturbative QCD
calculation for the gluon--$c$ quark scattering.  None of these calculations are reliable close to $T_c$. In order to get some hints of the evolution of $2\pi T D_x$ around
the phase transition, we use an adaptation of the recent lattice-QCD results of Ref.~\cite{Banerjee:2011ra}. In that work the diffusion coefficient is extracted using
the quenched approximation, i.e. there are no thermal light quarks and only dynamical gluons are considered. According to the authors of Ref.~\cite{Banerjee:2011ra}, 
the adimensional number $2\pi TD_x$ can directly be taken for real QCD (with dynamical light quarks) when considered as a function of $T/T_c$.
For real QCD we fix the crossover temperature at $T_c=160$ MeV. This number is consistent with recent estimates of the crossover temperature in lattice QCD \cite{Bazavov:2011nk, Borsanyi:2010bp}.

In spite of the mentioned caveats for each computation, an impressive agreement among the three calculations can be seen. A clear minimum of $2\pi T D_x$ appears at the phase transition temperature.

We can also analyze the possibility of a minimum in $2\pi T D_x$ close to the phase transition in the case of an isentropic trajectory. Within our approach, we can
easily obtain the value of $2\pi T D_x$ in the hadronic phase. However, we are not aware of an analogous calculation within the quark-gluon plasma. The simplest estimate for the spatial diffusion coefficient at high temperatures
comes from perturbative QCD. We thus generalize the results in Ref.~\cite{Moore:2004tg} for the diffusion coefficient at nonzero chemical potential. Although the validity of this calculation at physical temperatures is questionable, it can give a qualitative understanding of the $\mu_B$ dependence.

Following closely the derivation of the diffusion coefficient in the static limit in Ref.~\cite{Moore:2004tg}, and using their scattering amplitudes for quarks and gluons coming from the $t$-channel gluon exchange, we find
the following expression for the momentum diffusion coefficient $\Gamma$:
\be \Gamma = \frac{8}{9\pi} \alpha^2_s \int_0^\infty dk k^2 \int_0^{2k} dq \frac{q^3}{\left[ q^2 + m^2_D (T,\mu_f) \right]^2}
 \left[ \sum_f^{N_f} \frac{e^{(k-\mu_f)/T}}{\left( e^{(k-\mu_f)/T}+1\right)^2} \left( 2 - \frac{q^2}{2k^2} \right) + 3 \frac{e^{k/T}}{\left(e^{k/T}-1\right)^2} \left( 2 -\frac{q^2}{k^2}+\frac{q^4}{4k^4} \right) \right]\ , \ee
with $\alpha_s^2=g^2/4\pi$ the running strong coupling and $N_f$ the number of light flavors. The first term inside brackets comes from the light-quark scattering and it carries 
the quark chemical potential $\mu_f$ in the Fermi-Dirac function. The second term comes from the gluon scattering (with $N_c=3$) and it contains the Bose-Einstein distribution.
Then, $D_x$ is calculated using Eqs.~(\ref{eq:einstein},\ref{eq:dx}):
\be D_x = \frac{T^2}{\Gamma} . \ee

We also include the effect of the finite quark chemical potential in the Debye mass~\cite{Kapusta:2006pm,LeBellac}:
\be m_D^2 = g^2 \left[ \left( 1 + \frac{N_f}{6} \right) T^2 + \sum_f^{N_f} \frac{\mu_f^2}{2 \pi^2} \right] \ . \ee

To extract the entropy per baryon we use the equation of state for an ideal gas of massless quarks and gluons at lowest
order in the perturbative expansion. The pressure for $N_f$ light flavors reads~\cite{Kapusta:2006pm}
\be P(T,\mu_f)= \frac{7\pi^2}{60} T^4 + \sum_f^{N_f} \left( \frac{\mu_f^2 T^2}{2} + \frac{\mu_f^4}{4 \pi^2} \right) \ , \ee 
where we take $N_f=3$ with $\mu_u=\mu_d=\mu_B/3$ and $\mu_s=0$.
The entropy per baryon is simply obtained as
\be \frac{s}{n_B} = \frac{  \left( \frac{\pa P}{\pa T} \right)_{\mu_B} }{ \left( \frac{\pa P}{\pa \mu_B} \right)_T} \ , \ee
which is fixed to the same characteristic values of the hadronic trajectories {\it viz.} $s/n_B=10,20,30$. 
\begin{figure}
\centering
\includegraphics[width=0.45\textwidth]{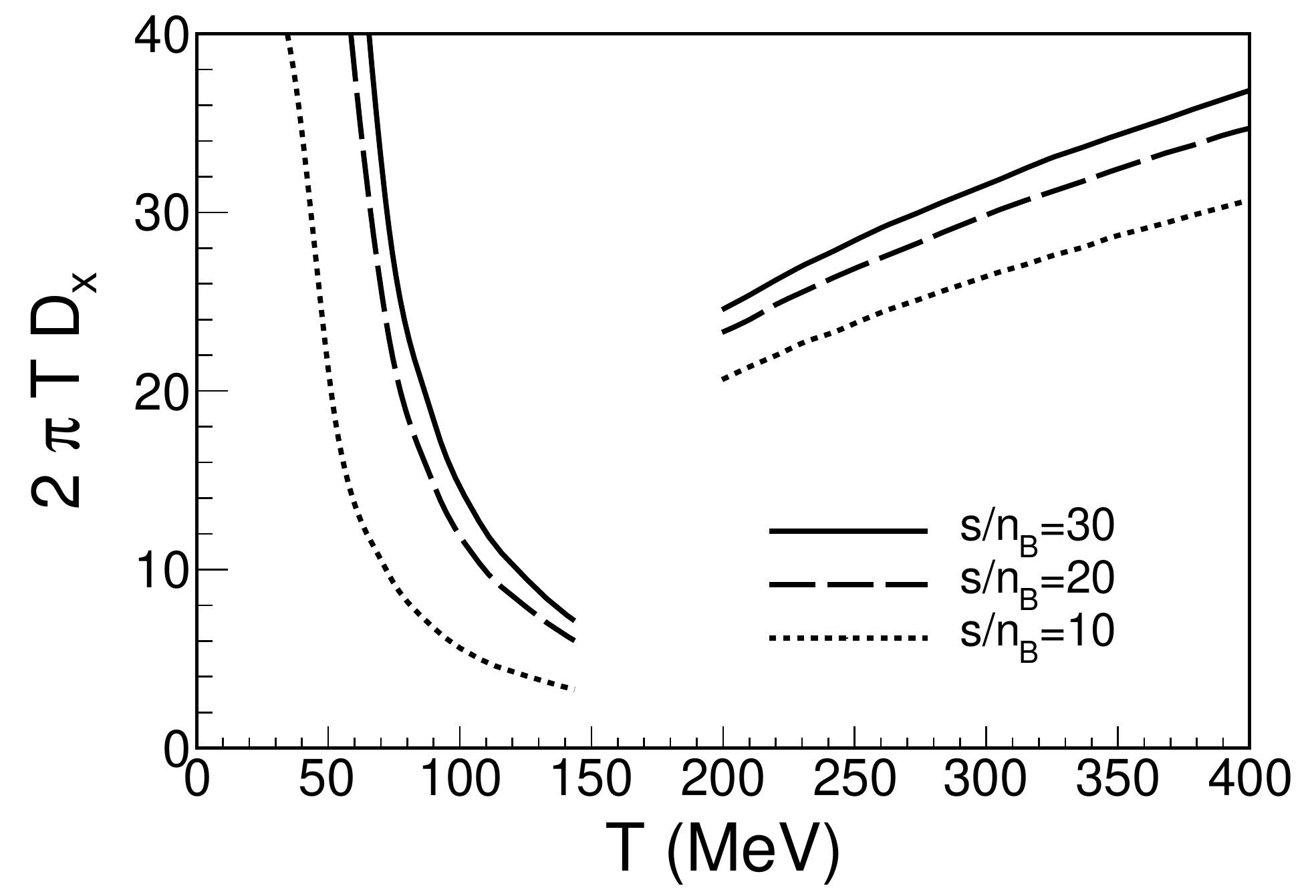}
\caption{\label{fig:transition2} Spatial diffusion coefficient for $D$ mesons multiplied by $2\pi T$ in the hadronic phase and the
perturbative quark-gluon plasma phase at finite chemical potential.}
\end{figure}

In Fig.~\ref{fig:transition2} we show the coefficient $2\pi T D_x$ around the transition temperature for the
hadronic gas and the perturbative QCD calculation. We observe that the dependence of the $2\pi T D_x$ on the entropy per baryon is similar in both phases: As long as
one increases the entropy per baryon (higher beam energies), the coefficient augments in both phases. The possible matching between curves in both phases for a
given $s/n_B$  seems to indicate the possible existence of a minimum in the $2\pi T D_x$ at the phase transition. However we obtain that the transition is not continuous.
We might not expect a smooth crossover between both phases but a first-order transition, where the transport coefficients (and other thermodynamical quantities) present a
finite discontinuity at $T_c$. However, it is known that the perturbative approach overpredicts this coefficient above $T_c$ at vanishing chemical potential by almost
an order of magnitude \cite{Rapp:2008qc,Rapp:2009my,Banerjee:2011ra}. Thus, a nonperturbative calculation for the quark-gluon-plasma diffusion coefficient at finite chemical
potential, in line with the one performed for vanishing chemical potential in Refs.~\cite{Rapp:2008qc,Rapp:2009my} would be most welcome. Work along this line is in progress.

\section{Conclusions and Outlook\label{sec:conclusions}}

We show our results for the $D$-meson drag and diffusion coefficients in a hot dense medium. We make use of a unitarized approach that takes, as a building block, an
effective-field-theory description of the interaction of $D$ mesons with light mesons, nucleons and $\Delta$'s, which is compatible with chiral and heavy quark
symmetries. We provide results in two different regimes. On one hand, we analyze drag and diffusion coefficients for vanishing baryochemical potential, $\mu_B=0$, which
is relevant for high beam energies like those reached at the LHC or RHIC. On the other hand, we study isentropic trajectories in the QCD phase diagram, expected at
heavy-ion collision experiments at FAIR and NICA. 
   
At $\mu_B=0$ the main contribution to the drag and diffusion coefficients comes from the interaction of $D$ mesons with pions as the thermal bath is mainly populated
by them, a fact which was already seen in Refs.~\cite{Ghosh:2011bw,He:2011yi,Abreu:2011ic}. Compared to previous works, our results are in good agreement with
the estimate in Ref.~\cite{He:2011yi} where the authors include effectively the most dominant resonant states in the meson and baryon sectors. However, our outcome
differs greatly from the results of Ref.~\cite{Ghosh:2011bw}, mostly due to the lack of unitarization in that reference.
   
Once we fix an isentropic trajectory, we find that the inclusion of $N$ and $\Delta$ baryons produces sizable effects in the $D$-meson drag force and diffusion
coefficients as compared to the $\mu_B=0$ case. This is due to the fact that the net abundance of baryons augments with increasing baryochemical potential. Thus, for
the lowest beam energies at FAIR, we find that at $T=140$ MeV the drag force increases by a factor of 2--3 with respect to its value at the same temperature for LHC
energies. In fact, we observe that the relaxation time grows with decreasing temperature, as expected from the behavior of the drag force coefficient. The comparison
between the obtained relaxation times and the typical fireball-expansion scale in heavy-ion collisions indicates that $D$ mesons are only partially equilibrated in the
hadronic phase. This can be tested, for example, in the future heavy-ion experiment CBM at FAIR~\cite{Friman:2011zz}.

We also present results for the spatial diffusion coefficient of a $D$ meson in hadronic matter and the evolution to the quark-gluon phase at zero and finite
baryochemical potential.  We analyze the plausibility of a minimum at the phase transition as an isentropic trajectory from the hadronic to the quark-gluon phase
is followed, in a similar manner as for the $\mu_B=0$ case \cite{Rapp:2008qc,Rapp:2009my}. Further studies, in particular in the quark-gluon domain, are needed in order to
verify our claim.

The transport coefficients obtained here are key parameters for the understanding of charm suppression and diffusion in the hadronic medium. Characteristic
observables are the nuclear suppression factor $R_{AA}$ and the elliptic flow $v_2$. These have been obtained from Langevin calculations~\cite{Rapp:2008qc,He:2012df,Lang:2012cx}
in both quark-gluon plasma and hadronic phases. Our results for $F(p)$ and $\Gamma_0(p),\Gamma_1(p)$ as a function of temperature and chemical potential can be well
accommodated to such approaches in order to improve the dissipative description in the hadronic evolution. Both observables have recently been extracted
in the ALICE experiment at LHC~\cite{ALICE:2012ab,delValle:2012qw}. Recent predictions for FAIR physics using the Langevin approach have been presented
in Ref.~\cite{Lang:2013cca}. It would be interesting to check whether the pure hadronic medium effects are as significant for FAIR physics as we describe in this work.

\begin{acknowledgments}
  We would like to thank L. Abreu for providing us with the coefficients of Table~\ref{tab:isoscoeff}, D. Banerjee for sharing with us the data for the diffusion coefficient
obtained from lattice QCD in Ref.~\cite{Banerjee:2011ra} and M. He for his parametrization of the $DN$ resonances from Ref.~\cite{He:2011yi}. We also acknowledge useful discussions with 
F. Llanes-Estrada, D. Cabrera, H. van Hees, T. Lang and R. Rapp. This work has been funded by Grants No. FPA2010-16963 (Ministerio de Ciencia e Innovaci\'on) and No. FP7-PEOPLE-2011-CIG under Contract
No. PCIG09-GA-2011-291679. L.T. acknowledges support from the Ram\'on y Cajal Research Programme (Ministerio de Ciencia e Innovaci\'on)
\end{acknowledgments}


\vspace{1cm}

\begin{thebibliography}{9}


\bibitem{Adare:2008ab} 
  A.~Adare {\it et al.}  [PHENIX Collaboration],
  Phys.\ Rev.\ Lett.\  {\bf 104}, 132301 (2010)
  [arXiv:0804.4168 [nucl-ex]].


\bibitem{Wilde:2012wc} 
  M.~Wilde [ALICE Collaboration],
  Nucl.\ Phys.\ A904-905 {\bf 2013}, 573c (2013)
  [arXiv:1210.5958 [hep-ex]].




\bibitem{Adcox:2001jp} 
  K.~Adcox {\it et al.}  [PHENIX Collaboration],
  Phys.\ Rev.\ Lett.\  {\bf 88}, 022301 (2002)
  [nucl-ex/0109003].


\bibitem{Aamodt:2011mr} 
  K.~Aamodt {\it et al.}  [ALICE Collaboration],
  Phys.\ Lett.\ B {\bf 696}, 328 (2011)
  [arXiv:1012.4035 [nucl-ex]].



\bibitem{Adare:2006nq} 
  A.~Adare {\it et al.}  [PHENIX Collaboration],
  Phys.\ Rev.\ Lett.\  {\bf 98}, 172301 (2007)
  [nucl-ex/0611018].


\bibitem{Masciocchi:2011fu} 
  S.~Masciocchi [ALICE Collaboration],
  J.\ Phys.\ G {\bf 38}, 124069 (2011)
  [arXiv:1109.6436 [nucl-ex]].


\bibitem{ALICE:2012ab} 
  B.~Abelev {\it et al.}  [ALICE Collaboration],
  J. High Energy Phys. {\bf 09} (2012) 112
 [arXiv:1203.2160 [nucl-ex]].


\bibitem{delValle:2012qw} 
  Z.~Conesa del Valle [ALICE Collaboration],
  Nucl.\ Phys.\ A904-905 {\bf 2013}, 178c (2013)
  [arXiv:1212.0385 [nucl-ex]].

\bibitem{Laine:2011is} 
  M.~Laine,
  J. High energy Phys. {\bf 04} (2011) 124
  [arXiv:1103.0372 [hep-ph]].


\bibitem{He:2011yi} 
  M.~He, R.~J.~Fries and R.~Rapp,
  Phys.\ Lett.\ B {\bf 701}, 445 (2011)
  [arXiv:1103.6279 [nucl-th]].


\bibitem{Ghosh:2011bw} 
  S.~Ghosh, S.~K.~Das, S.~Sarkar and J.~-eAlam,
  Phys.\ Rev.\ D {\bf 84}, 011503 (2011)
  [arXiv:1104.0163 [nucl-th]].



\bibitem{Abreu:2011ic} 
  L.~M.~Abreu, D.~Cabrera, F.~J.~Llanes-Estrada and J.~M.~Torres-Rincon,
  Ann. Phys.\  {\bf 326}, 2737 (2011)
  [arXiv:1104.3815 [hep-ph]].

\bibitem{Abreu:2012et} 
  L.~M.~Abreu, D.~Cabrera and J.~M.~Torres-Rincon,
  Phys.\ Rev.\ D {\bf 87}, 034019 (2013)
  [arXiv:1211.1331 [hep-ph]].

\bibitem{landau1981course}
  L.D.~Landau and E.M.~Lifshitz and L.P.~Pitaevskii, 
  \emph{Course of Theoretical Physics. vol. 10: Physical Kinetics},
  (Pergamon Press, 1981)



\bibitem{Mizutani:2006vq} 
  T.~Mizutani and A.~Ramos,
  Phys.\ Rev.\ C {\bf 74}, 065201 (2006)
  [hep-ph/0607257].


\bibitem{GarciaRecio:2008dp} 
  C.~Garcia-Recio, V.~K.~Magas, T.~Mizutani, J.~Nieves, A.~Ramos, L.~L.~Salcedo and L.~Tolos,
  Phys.\ Rev.\ D {\bf 79}, 054004 (2009)
  [arXiv:0807.2969 [hep-ph]].


\bibitem{Romanets:2012hm} 
  O.~Romanets, L.~Tolos, C.~Garcia-Recio, J.~Nieves, L.~L.~Salcedo and R.~G.~E.~Timmermans,
  Phys.\ Rev.\ D {\bf 85}, 114032 (2012).




\bibitem{Oller:1997ti}
  J.~A.~Oller and E.~Oset,
  Nucl.\ Phys.\  A {\bf 620 } (1997)  438-456.
  [hep-ph/9702314].


\bibitem{Oller:2000fj} 
  J.~A.~Oller and U.~G.~Meissner,
  Phys.\ Lett.\ B {\bf 500}, 263 (2001)
  [hep-ph/0011146].


\bibitem{Kolomeitsev:2003ac} 
  E.~E.~Kolomeitsev and M.~F.~M.~Lutz,
  Phys.\ Lett.\ B {\bf 582}, 39 (2004)
  [hep-ph/0307133].
  
\bibitem{Hofmann:2003je} 
  J.~Hofmann and M.~F.~M.~Lutz,
  Nucl.\ Phys.\ A {\bf 733}, 142 (2004)
  [hep-ph/0308263].
  
  
\bibitem{Guo:2006fu} 
  F.~-K.~Guo, P.~-N.~Shen, H.~-C.~Chiang, R.~-G.~Ping and B.~-S.~Zou,
  Phys.\ Lett.\ B {\bf 641}, 278 (2006)
  [hep-ph/0603072].
  
\bibitem{dany} 
  D.~Gamermann, E.~Oset, D.~Strottman and M.~J.~Vicente Vacas,
  Phys.\ Rev.\ D {\bf 76}, 074016 (2007)
  [hep-ph/0612179].
  
\bibitem{Molina:2008nh} 
  R.~Molina, D.~Gamermann, E.~Oset and L.~Tolos,
  Eur.\ Phys.\ J.\ A {\bf 42}, 31 (2009)
  [arXiv:0806.3711 [nucl-th]].


\bibitem{danyax} 
  D.~Gamermann and E.~Oset,
  Eur.\ Phys.\ J.\ A {\bf 33}, 119 (2007)
  [arXiv:0704.2314 [hep-ph]].
  
  
\bibitem{Branz:2009yt} 
  T.~Branz, T.~Gutsche and V.~E.~Lyubovitskij,
  Phys.\ Rev.\ D {\bf 80}, 054019 (2009)
  [arXiv:0903.5424 [hep-ph]].
  
\bibitem{Faessler:2007gv} 
  A.~Faessler, T.~Gutsche, V.~E.~Lyubovitskij and Y.~-L.~Ma,
  Phys.\ Rev.\ D {\bf 76}, 014005 (2007)
  [arXiv:0705.0254 [hep-ph]].
  
\bibitem{Segovia:2008zz} 
  J.~Segovia, A.~M.~Yasser, D.~R.~Entem and F.~Fernandez,
  Phys.\ Rev.\ D {\bf 78}, 114033 (2008).
  
\bibitem{FernandezCarames:2009zz} 
  T.~Fernandez-Carames, A.~Valcarce and J.~Vijande,
  Phys.\ Rev.\ Lett.\  {\bf 103}, 222001 (2009)
  [arXiv:1001.4506 [hep-ph]].
  
\bibitem{Gutsche:2010zza} 
  T.~Gutsche and V.~E.~Lyubovitskij,
  AIP Conf.\ Proc.\  {\bf 1257}, 385 (2010).
  
  
\bibitem{HidalgoDuque:2012pq} 
  C.~Hidalgo-Duque, J.~Nieves and M.~P.~Valderrama,
  Phys.\  Rev.\  D 87, {\bf 076006} (2013)
  [arXiv:1210.5431 [hep-ph]].
  
  
  
\bibitem{Guo:2008zg} 
  F.~-K.~Guo, C.~Hanhart and U.~-G.~Meissner,
  Phys.\ Lett.\ B {\bf 665}, 26 (2008)
  [arXiv:0803.1392 [hep-ph]].


\bibitem{Lutz:2007sk} 
  M.~F.~M.~Lutz and M.~Soyeur,
  Nucl.\ Phys.\ A {\bf 813}, 14 (2008)
  [arXiv:0710.1545 [hep-ph]].
  

\bibitem{Guo:2008gp} 
  F.~-K.~Guo, C.~Hanhart, S.~Krewald and U.~-G.~Meissner,
  Phys.\ Lett.\ B {\bf 666}, 251 (2008)
  [arXiv:0806.3374 [hep-ph]].



\bibitem{Guo:2009ct} 
  F.~-K.~Guo, C.~Hanhart and U.~-G.~Meissner,
  Eur.\ Phys.\ J.\ A {\bf 40}, 171 (2009)
  [arXiv:0901.1597 [hep-ph]].
  
\bibitem{Geng:2010vw} 
  L.~S.~Geng, N.~Kaiser, J.~Martin-Camalich and W.~Weise,
  Phys.\ Rev.\ D {\bf 82}, 054022 (2010)
  [arXiv:1008.0383 [hep-ph]].


\bibitem{Liu:2012zya} 
  L.~Liu, K.~Orginos, F.~-K.~Guo, C.~Hanhart and U.~-G.~Meissner,
  Phys.\ Rev.\ D {\bf 87}, 014508 (2013)
  [arXiv:1208.4535 [hep-lat]].


 \bibitem{Tolos:2004yg}
  L.~Tolos, J.~Schaffner-Bielich and A.~Mishra,
  Phys.\ Rev.\  C {\bf 70}, 025203 (2004).

 
\bibitem{Tolos:2005ft} 
  L.~Tolos, J.~Schaffner-Bielich and H.~Stoecker,
  Phys.\ Lett.\ B {\bf 635}, 85 (2006).


\bibitem{Lutz:2003jw}
  M.~F.~M.~Lutz and E.~E.~Kolomeitsev,
  Nucl.\ Phys.\  A {\bf 730}, 110 (2004).


\bibitem{Lutz:2005ip} 
  M.~F.~M.~Lutz and E.~E.~Kolomeitsev,
  Nucl.\ Phys.\ A {\bf 755}, 29 (2005)



\bibitem{Hofmann:2005sw} 
  J.~Hofmann and M.~F.~M.~Lutz,
  Nucl.\ Phys.\ A {\bf 763}, 90 (2005).


\bibitem{Hofmann:2006qx}
  J.~Hofmann and M.~F.~M.~Lutz,
  Nucl.\ Phys.\  A {\bf 776}, 17 (2006).

\bibitem{Lutz:2005vx} 
  M.~F.~M.~Lutz and C.~L.~Korpa,
  Phys.\ Lett.\ B {\bf 633}, 43 (2006).

\bibitem{Tolos:2007vh}
  L.~Tolos, A.~Ramos and T.~Mizutani,
  Phys.\ Rev.\  C {\bf 77}, 015207 (2008).


\bibitem{JimenezTejero:2009vq} 
  C.~E.~Jimenez-Tejero, A.~Ramos and I.~Vidana,
  Phys.\ Rev.\ C {\bf 80}, 055206 (2009).


\bibitem{Haidenbauer:2007jq}
  J.~Haidenbauer, G.~Krein, U.~G.~Meissner and A.~Sibirtsev,
  Eur.\ Phys.\ J.\  A {\bf 33}, 107 (2007).

\bibitem{Haidenbauer:2008ff}
  J.~Haidenbauer, G.~Krein, U.~G.~Meissner and A.~Sibirtsev,
  Eur.\ Phys.\ J.\  A {\bf 37}, 55 (2008).

\bibitem{Haidenbauer:2010ch}
  J.~Haidenbauer, G.~Krein, U.~G.~Meissner and L.~Tolos,
  Eur. Phys. J A {\bf 47}, 18 (2011).

\bibitem{Wu:2010jy}
  J.~-J.~Wu, R.~Molina, E.~Oset and B.~S.~Zou,
  Phys.\ Rev.\ Lett.\  {\bf 105}, 232001 (2010).
 
\bibitem{Wu:2010vk} 
  J.~-J.~Wu, R.~Molina, E.~Oset and B.~S.~Zou,
  Phys.\ Rev.\ C {\bf 84}, 015202 (2011).
 
\bibitem{Wu:2012md} 
  J.~-J.~Wu, T.~-S.~H.~Lee and B.~S.~Zou,
  Phys.\ Rev.\ C {\bf 85}, 044002 (2012).

\bibitem{Oset:2012ap} 
  E.~Oset, A.~Ramos, E.~J.~Garzon, R.~Molina, L.~Tolos, C.~W.~Xiao, J.~J.~Wu and
B.~S.~Zou,
  Int.\ J.\ Mod.\ Phys.\ E {\bf 21}, 1230011 (2012).

\bibitem{Garcia-Recio:2013gaa} 
  C.~Garcia-Recio, J.~Nieves, O.~Romanets, L.~L.~Salcedo and L.~Tolos,
  Phys.\ Rev.\ D {\bf 87}, 074034 (2013)
  [arXiv:1302.6938 [hep-ph]].


\bibitem{Gamermann:2011mq} 
  D.~Gamermann, C.~Garcia-Recio, J.~Nieves and L.~L.~Salcedo,
  Phys.\ Rev.\ D {\bf 84}, 056017 (2011).



  

\bibitem{Nieves:2001wt} 
  J.~Nieves and E.~Ruiz Arriola,
  Phys.\ Rev.\ D {\bf 64}, 116008 (2001).

\bibitem{Xiao:2013jla}
  C.~W.~Xiao and E.~Oset,
  arXiv:1305.0786 [hep-ph].
  
\bibitem{fick}
  A.~Fick,
  Ann. der Physik 170, 59 (1855)


\bibitem{smith1989transport}
  H.~Smith, and H.H.~Jensen,
  \emph{Transport Phenomena},
  (Oxford University Press, USA. 1989)

\bibitem{chapman1952mathematical}
  S.~Chapman, and T.G.~Cowling,   
 \emph{The Mathematical Theory of Mon-uniform gases},
 (Cambridge University Press. 1970)

\bibitem{Das:2011vba} 
  S.~K.~Das, S.~Ghosh, S.~Sarkar and J.~-eAlam,
  Phys.\ Rev.\ D {\bf 85}, 074017 (2012)
  [arXiv:1109.3359 [hep-ph]].

\bibitem{Bravina:2008ra} 
  L.~V.~Bravina, I.~Arsene, M.~S.~Nilsson, K.~Tywoniuk, E.~E.~Zabrodin, J.~Bleibel, A.~Faessler and C.~Fuchs {\it et al.},
  Phys.\ Rev.\ C {\bf 78}, 014907 (2008)
  [arXiv:0804.1484 [hep-ph]].


  
\bibitem{Csernai:2006zz} 
  L.~P.~Csernai, J.~.I.~Kapusta and L.~D.~McLerran,
  Phys.\ Rev.\ Lett.\  {\bf 97}, 152303 (2006)
  [nucl-th/0604032].

\bibitem{Dobado:2008vt} 
  A.~Dobado, F.~J.~Llanes-Estrada and J.~M.~Torres-Rincon,
  Phys.\ Rev.\ D {\bf 79}, 014002 (2009)
  [arXiv:0803.3275 [hep-ph]].

\bibitem{Arnold:2003zc} 
  P.~B.~Arnold, G.~D.~Moore and L.~G.~Yaffe,
  J. High Energy PHys. {\bf 05}, 051 (2003)
  [hep-ph/0302165].


\bibitem{Schafer:2009dj} 
  T.~Schäfer and D.~Teaney,
  Rep.\ Prog.\ Phys.\  {\bf 72}, 126001 (2009)
  [arXiv:0904.3107 [hep-ph]].


\bibitem{Chakraborty:2010fr} 
  P.~Chakraborty and J.~I.~Kapusta,
  Phys.\ Rev.\ C {\bf 83}, 014906 (2011)
  [arXiv:1006.0257 [nucl-th]].
  
    
  
\bibitem{Dobado:2009ek} 
  A.~Dobado, F.~J.~Llanes-Estrada and J.~M.~Torres-Rincon,
  Phys.\ Rev.\ D {\bf 80}, 114015 (2009)
  [arXiv:0907.5483 [hep-ph]].






\bibitem{Sasaki:2008um} 
  C.~Sasaki and K.~Redlich,
  Nucl.\ Phys.\ A {\bf 832}, 62 (2010)
  [arXiv:0811.4708 [hep-ph]].

\bibitem{Khvorostukhin:2010aj} 
  A.~S.~Khvorostukhin, V.~D.~Toneev and D.~N.~Voskresensky,
  Nucl.\ Phys.\ A {\bf 845}, 106 (2010)
  [arXiv:1003.3531 [nucl-th]].

  
\bibitem{Dobado:2012zf} 
  A.~Dobado and J.~M.~Torres-Rincon,
  Phys.\ Rev.\ D {\bf 86}, 074021 (2012)
  [arXiv:1206.1261 [hep-ph]].
  


\bibitem{Kharzeev:2007wb} 
  D.~Kharzeev and K.~Tuchin,
  J. High Energy Phys. {\bf 09} (2008) 093
  [arXiv:0705.4280 [hep-ph]].

\bibitem{Karsch:2007jc} 
  F.~Karsch, D.~Kharzeev and K.~Tuchin,
  Phys.\ Lett.\ B {\bf 663}, 217 (2008)
  [arXiv:0711.0914 [hep-ph]].


   
\bibitem{Rapp:2008qc} 
  R.~Rapp and H.~van Hees,
  arXiv:0803.0901 [hep-ph].
  

  
  
\bibitem{Banerjee:2011ra} 
  D.~Banerjee, S.~Datta, R.~Gavai and P.~Majumdar,
  Phys.\ Rev.\ D {\bf 85}, 014510 (2012)
  [arXiv:1109.5738 [hep-lat]].
  
\bibitem{Bazavov:2011nk} 
  A.~Bazavov, T.~Bhattacharya, M.~Cheng, C.~DeTar, H.~T.~Ding, S.~Gottlieb, R.~Gupta and P.~Hegde {\it et al.},
  Phys.\ Rev.\ D {\bf 85}, 054503 (2012)
  [arXiv:1111.1710 [hep-lat]].
  

\bibitem{Borsanyi:2010bp} 
  S.~Borsanyi {\it et al.}  [Wuppertal-Budapest Collaboration],
  J. High Energy Phys. {\bf 09} (2010) 073
  [arXiv:1005.3508 [hep-lat]].
  
  
\bibitem{Moore:2004tg} 
  G.~D.~Moore and D.~Teaney,
  Phys.\ Rev.\ C {\bf 71}, 064904 (2005)
  [hep-ph/0412346].
  
  
\bibitem{Kapusta:2006pm} 
  J.~I.~Kapusta and C.~Gale,
  \emph{Finite-temperature field theory: Principles and Applications},
  (Cambridge University Press, 2006)


\bibitem{LeBellac}
  M.~Le Bellac  
  \emph{Thermal Field Theory},
  (Cambridge University Press, 2000)
  
\bibitem{Rapp:2009my} 
  R.~Rapp and H.~van Hees,
  in R. C. Hwa, X.-N. Wang (Ed.) Quark Gluon Plasma 4, World Scientific, 111 (2010)
  [arXiv:0903.1096 [hep-ph]].


\bibitem{Friman:2011zz} 
  B.~Friman, C.~Hohne, J.~Knoll, S.~Leupold, J.~Randrup, R.~Rapp and P.~Senger,
  Lect.\ Notes Phys.\  {\bf 814}, 1 (2011).

  
\bibitem{He:2012df} 
  M.~He, R.~J.~Fries and R.~Rapp,
  Phys.\ Rev.\ Lett.\  {\bf 110}, 112301 (2013)
  [arXiv:1204.4442 [nucl-th]].
  
\bibitem{Lang:2012cx} 
  T.~Lang, H.~van Hees, J.~Steinheimer and M.~Bleicher,
  arXiv:1211.6912 [hep-ph].


\bibitem{Lang:2013cca} 
  T.~Lang, H.~van Hees, J.~Steinheimer and M.~Bleicher,
  arXiv:1305.1797 [hep-ph].



\end{thebibliography}
\end{document}